\documentclass[preprint,prb,superscriptaddress,showpacs]{revtex4-1}%
\usepackage{amssymb, amsmath}
\usepackage{float}
\usepackage{graphicx}
\usepackage{color}
\usepackage{amsmath}
\usepackage{amsfonts}
\usepackage{amssymb}%
\providecommand{\U}[1]{\protect\rule{.1in}{.1in}}
\makeatletter
\renewcommand*{\fnum@figure}{{\normalfont\bfseries \figurename~\thefigure}}
\renewcommand*{\@caption@fignum@sep}{\textbf{ : }}
\makeatother
\usepackage{hyperref}
\hypersetup{
    colorlinks=true,
    linkcolor=blue,
    filecolor=magenta,      
    urlcolor=cyan,
}

\begin{document}
\title{Novel magnetic states and nematic spin chirality in the kagome lattice metal
YMn$_{6}$Sn$_{6}$}

\author{Nirmal J. Ghimire}
\thanks{corresponding author}
\email{nghimire@gmu.edu}
\affiliation{Department of Physics and Astronomy, George Mason University, Fairfax, VA 22030}
\affiliation{Quantum Science and Engineering Center, George Mason University, Fairfax, VA 22030}
\author{Rebecca L. Dally}
\affiliation{NIST Center for Neutron Research, National Institute of Standards and
Technology, Gaithersburg, MD 20899}
\author{L. Poudel}
\affiliation{NIST Center for Neutron Research, National Institute of Standards and
Technology, Gaithersburg, MD 20899}
\affiliation{Department of Materials Science and Engineering, University of Maryland,
College Park, MD 20742}
\author{D. C. Jones}
\affiliation{Department of Physics and Astronomy, George Mason University, Fairfax, VA 22030}
\affiliation{Quantum Science and Engineering Center, George Mason University, Fairfax, VA 22030}
\author{D. Michel}
\affiliation{Department of Physics and Astronomy, George Mason University, Fairfax, VA 22030}
\affiliation{Quantum Science and Engineering Center, George Mason University, Fairfax, VA 22030}
\author{N. Thapa Magar}
\affiliation{Department of Physics and Astronomy, George Mason University, Fairfax, VA 22030}
\author{M. Bleuel}
\affiliation{NIST Center for Neutron Research, National Institute of Standards and
Technology, Gaithersburg, MD 20899}
\affiliation{Department of Materials Science and Engineering, University of Maryland,
College Park, MD 20742}
\author{Michael A. McGuire}
\affiliation{Materials Science and Technology Division, Oak Ridge National Laboratory, Oak
Ridge, Tennessee 37831, United States}
\author{J. S. Jiang}
\affiliation{Materials Science Division, Argonne National Laboratory, 9700 South Cass
Avenue, Argonne, Illinois 60439, United States}
\author{J. F. Mitchell}
\affiliation{Materials Science Division, Argonne National Laboratory, 9700 South Cass
Avenue, Argonne, Illinois 60439, United States}
\author{Jeffrey W. Lynn}
\affiliation{NIST Center for Neutron Research, National Institute of Standards and
Technology, Gaithersburg, MD 20899}
\author{I. I. Mazin}
\affiliation{Department of Physics and Astronomy, George Mason University, Fairfax, VA 22030}
\affiliation{Quantum Science and Engineering Center, George Mason University, Fairfax, VA 22030}

\date{\today}
\maketitle
\textbf{Identification, understanding, and manipulation of novel magnetic
textures is essential for the discovery of new quantum materials for
future spin-based electronic devices. In particular, materials that
manifest a large response to external stimuli such as a magnetic field are
subject to intense investigation. Here, we study the kagome-net magnet
YMn$_{6}$Sn$_{6}$ by magnetometry, transport, and neutron diffraction
measurements combined with first principles calculations. We identify a number
of nontrivial magnetic phases, explain their microscopic nature, and demonstrate that one of them hosts a large
topological Hall effect (THE). We propose a new nematic chirality mechanism,
reminiscent of the nematicity in Fe-based superconductors, which leads to the
THE at elevated temperatures. This
interesting physics comes from parametrically frustrated interplanar exchange
interactions that trigger strong magnetic fluctuations. Our results pave a
path to new chiral spin textures, promising for novel spintronics.}
\\

Kagome planes formed by Fe or Mn often have strong in-plane ferromagnetic (FM)
exchange interactions which are not magnetically frustrated but still have
 features typical of kagome lattices - Dirac and flat bands - providing an ideal platform for novel topological states 
\cite{Ghimire2020,Kang2020d,Ye2018c,Liu2018a,Yin2019b,Zhang2011e}. The
interplanar interactions, on the other hand, are much weaker and often
frustrated. FM ordering in the two-dimensional planes is then strongly
suppressed due to the Mermin-Wagner theorem, enabling very strong magnetic
fluctuations at elevated temperatures that provide fertile ground for new and
interesting phenomena \cite{Grohol2005,Pereiro2014,Hirschberger2019,Yin2018a}.

YMn$_{6}$Sn$_{6}$ is a prototype for this materials class. It forms an
hexagonal P6/mmm structure ($a=5.540$ \AA \ and $c$ = 9.020 \AA ) consisting of kagome planes [Mn$_{3}$Sn] separated by two inequivalent
Sn$_{3},$ and Sn$_{2}$Y layers, $i.e.,$ [Mn$_{3}$Sn][Sn$_{3}]$[Mn$_{3}$%
Sn]$[$Sn$_{2}$Y] [Figs. \ref{Fig1}(a),(b)]. YMn$_{6}$Sn$_{6}$ is a good metal
[Fig. \ref{Fig1}(c)], and as such is expected to have relatively long-range
exchange interactions, possibly including Ruderman-Kittel-Kasuya-Yosida (RKKY)
coupling [Fig. \ref{Fig1}(a)]. All Mn planes and in-plane nearest neighbor
Mn-Mn bonds are crystallographically equivalent, but the interplanar Mn-Mn
bonds along $c$ are dramatically different, with a FM exchange interaction
across the Sn$_{3}$ layers, and antiferromagnetic (AF) across the Sn$_{2}$Y layers. These are frustrated by the \textit{second} neighbor interaction across an
intermediate Mn$_{3}$Sn layer ($J_{1}$ and $J_{3}$ are FM, while $J_{2}$ is
AF) and result in complex magnetic behaviors \cite{Venturini1996a,Uhlirova2006}. Below $T_{N}$ $\approx$ 345 K [Fig.
\ref{Fig1}(d)], a commensurate collinear AF structure forms with the
propagation vector \textbf{k} = (0, 0, 0.5). On cooling, an incommensurate
phase quickly appears, which coexists with the commensurate phase in a narrow
temperature range and becomes the only phase below 300 K
\cite{Venturini1996a,Zhang2020}. Based on powder diffraction, the incommensurate state has been reported to
have two (and even three at room temperature) nearly equal wave
vectors,\cite{Venturini1996a} which can be described as a staggered spiral,
also dubbed the \textquotedblleft double flat spiral," \cite{Rosenfeld2008a} as
depicted in Fig. \ref{Fig1}(e). A magnetic field applied in the $ab$-plane
induces multiple transitions seen in the magnetization and Hall resistivity
\cite{Uhlirova2006}. Interestingly, an enigmatic topological Hall effect (THE)
is observed at elevated temperatures, with the largest value around 245 K and
a magnetic field of 4 T \cite{Wang2019a}. In this article, we determine the
microscopic origin of the magnetic field-induced phases of YMn$_{6}$Sn$_{6}$
and develop a theory describing the observed THE.

We first map out the different field-induced magnetic phases
of YMn$_{6}$Sn$_{6}$ with bulk measurements. Figure \ref{Fig2}(a) shows the
magnetization measurements of YMn$_{6}$Sn$_{6}$ at two representative
temperatures, 5 K and 245 K. For the magnetic field applied along the $c$-axis
(red curve) the magnetization increases smoothly with field and for 5 K
saturates slightly above 12 T, while the 245 K data show that the saturation
field clearly decreases with increasing temperature. The effect of a magnetic
field applied in the $ab$-plane ($H_{ab}$) shown by the blue curves is more
dramatic. At 5 K we see a sharp increase at 2 T indicative of a metamagnetic
transition. A closer look reveals two close transitions, more apparent in the
ac-susceptibility measurement [Fig.~\ref{Fig2}(b)]. Since the two transitions
are very close, we denote the metamagnetic transition field by a single
variable, $H_{1}$, for the remainder of the paper. As the field is further
increased, the magnetization changes slope and increases continuously until
$H_{2}$ = 7 T. Above $H_{2}$, the magnetization grows slower, and saturates at
$H_{3}$ = 9.8 T. As temperature is increased, $H_{1}$, $H_{2}$ and $H_{3}$ all
shift to lower fields, and $H_{2}$ and $H_{3}$ become closer and merge. A
phase diagram constructed from the ac-susceptibility is depicted in Fig.
\ref{Fig2}(b), with four main phases: (1) $0<H<H_{1}$, (2) $H_{1}<H<H_{2}$,
(3) $H_{2}<H<H_{3}$ and (4) $H>H_{3}$. We call them distorted spiral (DS),
transverse conical spiral (TCS), fan-like (FL), and forced-ferromagnetic (FF),
respectively, based on the magnetic structures as detailed below. The narrow
intermediate phases between FL and FF, and between TCS and FF are labelled
\textquotedblleft I" and \textquotedblleft II", respectively.

The Hall resistivity ($\rho_{H}$) and magnetization ($M$) as a function of
$H_{ab}$ at 5 K and 245 K are compared in Figs. \ref{Fig2}(c) and (d),
respectively. At 5 K, $\rho_{H}$ has a very small negative slope in the DS
phase. At $H_{1}$, $\rho_{H}$ shows a small jump but then decreases before
increasing rapidly to saturation in the FF state, forming a remarkable minimum
in the FL phase. The behavior of $\rho_{H}$ is significantly different at 245
K, where it exhibits a positive slope in the DS phase. At the metamagnetic
transition ($H_{1}$), it shows a sizable jump, then increases non-linearly
with the magnetization in the TCS phase, which has been interpreted as the
topological Hall effect (THE)\cite{Wang2019a}.

The zero-field neutron diffraction data are plotted in Fig. \ref{Fig3}(a). A
commensurate magnetic Bragg peak is observed at the onset of long range
magnetic order, where \textbf{k} = (0, 0, 0.5) and $T_{N}$ = 345 K, which
quickly transforms into two distinct wave vectors. These two incommensurate
structures coexist from their onset to the base temperature (12 K) determined
by high resolution measurements [inset in Fig. \ref{Fig3}(a)]. The two wave
vectors (0, 0, $k_{z,1}$) and (0, 0, $k_{z,2}$) with $k_{z,1}$ $<$ $k_{z,2}$
evolve smoothly with temperature along $L$, and $|$$k_{z,1}-k_{z,2}$$|$
decreases with cooling. The two magnetic structures stemming from $k_{z,1}$
and $k_{z,2}$ are consistent with previous
reports,\cite{Venturini1996a,Rosenfeld2008a} [see Fig. \ref{Fig1}(e)] but with
slightly different periodicities [Fig. \ref{SFN2}(a)].

We now focus on the multiple magnetic phases induced via application of an
external magnetic field in the $ab$-plane. Figures \ref{Fig3}(b)-(c) show data
taken about $(0,0,2-k_{z,n})$ ($n=1,2$) for 100 K and 256 K, respectively. We
find that $k_{z,n}$ are almost field-independent, except for an abrupt shift
to larger momentum for both magnetic peaks at $H_{1},$ which lies between 2.0
T and 2.5 T (between 1.5 T and 2.0 T for $H_{1}$ at 256 K). Concomitant with
these shifts are pronounced decreases in intensity of the Bragg peaks at (0,
0, $L\pm k_{z,n}$) positions. The $T$ = 100 K data show a new commensurate
structure emerging at $H_{2}$ (6 T), with the wave vector $(0,0,k_{c}),$ where
$k_{c}$ = 0.25, plus a satellite at $2k_{c}$-type positions which can be seen
in Fig.~\ref{Fig3}(e) (discussed more below). These commensurate peaks coexist
with the incommensurate peaks at 6 T and emerge at the cost of the
incommensurate intensities [see Supplementary Information 1 (\ref{SI1}) for details].

A 15 T magnet was employed to focus on the high field behavior, where a coarse
instrumental resolution was used to compensate for the reduced intensities.
The two incommensurate wave vectors are not resolvable with this resolution,
but the data satisfactorily capture the overall high field behavior. Figure
\ref{Fig3}(f) shows that at 200 K the incommensurate peaks disappear above 7
T, similar to the observation that they are almost fully suppressed by 6 T in
the 256 K high resolution data [Fig.~\ref{Fig3}(c)]. The $\mathbf{k}%
=(0,0,0.25)$ commensurate structure at 100 K and 6 T in Fig.~\ref{Fig3}(b) can
be seen at the same field in Fig.~\ref{Fig3}(e) at $\mathbf{Q}=(0,0,2.25)$
with a satellite peak at $\mathbf{Q}=(0,0,2.50)$. Additionally, we see that
all but the FF structures disappear above 8 T as the spins become fully
polarized. The high-field commensurate phase persists down to 10 K, shown in
Fig.~\ref{Fig3}(d), but is shifted higher in field and is present between 6.5
T and 9.5 T.

The neutron data capture all the features observed in bulk magnetic
measurements (Fig. \ref{Fig2}). Below $H_{1}$ there is very little change to
the incommensurate peaks. At $H_{1}$, the wave vector positions change by
$\sim$3\% and intensity by up to 60\% [see Fig. \ref{SFN2}]. The $H_{1}$
transition, which spans almost the entire temperature range of zero-field
incommensurability, resembles a spin-flop transition, deduced from the
magnetization data in Fig. \ref{Fig2}(a). As discussed further in the
theoretical section, this is a spin flop from a helical to cycloidal spiral.
Comparison of the structure factor calculations for each magnetic structure to
the data supports this assignment (see \ref{SI1} and Fig. \ref{SFN3} for details). At $H_{2}$,
commensurate peaks with \textbf{k} = (0, 0, 0.25)-type positions, and
satellites at 2$k_{c}$, emerge at the cost of the incommensurate structures.
The commensurate peaks only appear in the FL phase. Curiously, however, a
commensurate phase with propagation vector \textbf{k} = (0, 0, 0.5) is seen to
emerge at 300 K and low field (2 T) as seen in Fig. \ref{Fig3}(g). This
temperature and field reside within region \textquotedblleft II" of the phase
diagram in Fig. \ref{Fig2}(b).

To understand the microscopic origin and nature of the different magnetic
phases, we performed first principles Density Functional Theory (DFT)
calculations and used the results to construct a mean field theory (MFT) at
$T$= 0. The details are presented in the methods and \ref{SI2}, and here we
summarize the main findings. First, DFT total energy calculations were
performed and fit to the Hamiltonian (Eqn. \ref{H})%

\begin{equation}
\mathcal{H}={\displaystyle\sum\limits_{i,j}}J_{n}\mathbf{n}_{i}\cdot
\mathbf{n}_{j}+{\displaystyle\sum\limits_{i,j}}J_{p}\mathbf{n}_{i}%
\cdot\mathbf{n}_{j}+K{\displaystyle\sum\limits_{i}}(n_{i}^{z})^{2}%
+{\displaystyle\sum\limits_{i}}J^{z}n_{i}^{z}\cdot n_{i+1}^{z}%
+{\displaystyle\sum\limits_{i}}\mathbf{n}_{i}\cdot\mathbf{H,}\label{H}%
\end{equation}
where $\mathbf{H}$ is the external field and $\mathbf{n}$ is a unit vector
along the local magnetization direction. The first sum runs over 6 nearest
neighbors along the $c$-axis, the second over the first neighbors in the
$ab$-plane, and the last three over all atoms ($i$+1 denotes the nearest
$c$-neighbor). $K$ is the easy-plane single-ion anisotropy, and the Ising-type anisotropic exchange, $J^{z}$, is the only one
allowed by symmetry for the vertical bonds. To account for Hubbard
correlations, we added a DFT+U correction (see \ref{SI2}). We found that the best
description of the ground state is attained for $U-J=0.4-0.6$ eV, and in the
following we use 0.4 (not unreasonable for a good metal). The results are
shown in the Supplementary Table \ref{T1} for three models: \textquotedblleft
full\textquotedblright, \textquotedblleft reduced\textquotedblright, where
$J_{4-6}$ are absorbed into modified $J_{2-3}$, and \textquotedblleft
minimal\textquotedblright, where $J_{z}$ is in addition combined with $K$. The
\textquotedblleft full\textquotedblright model has a staggered spiral as a
ground state, as shown in Fig. \ref{Fig1}(e), with the two angles
$\alpha=-22^{\circ}$ and $\beta=138^{\circ}$, in reasonable agreement with the
low temperature experimental \textbf{k} $\approx$ (0, 0, 0.25) described by
the pitching angles $-20^{\circ}$ and $\beta=110^{\circ}$ ($\alpha+\beta
=90{{}^{\circ}}$). These angles were used in the \textquotedblleft
reduced\textquotedblright  model, $J_{2-3}$ calculations.

We now present the MFT results for the \textquotedblleft
minimal\textquotedblright  and \textquotedblleft reduced\textquotedblright
 models at $T$ = 0. At $H$ = 0 one gets a staggered
spiral\cite{Rosenfeld2008a}. Without $K$, the minimal model uniquely defines
\cite{Rosenfeld2008a} the propagation vector $k_{z}$ but is degenerate with
respect to the plane in which the magnetic moments rotate. The anisotropy $K$
locks the spins to the $ab$-plane. Results for the MFT are presented in Figs.
\ref{Fig4}(a)-(c) (see \ref{SI3} for details). The behavior for $H||c$ is trivial:
the helical spiral becomes longitudinal conical spiral (LCS) and gradually
transforms into a field-polarized FM phase. For $H||a$, if there were no magnetic anisotropy
($K$= 0), the staggered spiral would immediately flop from spins rotating in
the $ab$-plane (helical) to those rotating in the $bc$-plane (cycloidal),
which would then gradually cant into a transverse conical spiral (TCS) state,
and eventually saturate. The magnetic anisotropy sets a finite spin-flop field
$H_{1}$ $\propto$ $\sqrt{\left\langle J\right\rangle K},$ where $\left\langle
J\right\rangle $ is the appropriately averaged $J_{1-3}$ parameters. Below
$H_{1}$, the spiral remains flat, but distorts slightly by canting each spin a
little toward $a$ (this is the DS phase). At $H_{1}$, the magnetization
increases discontinuously. However, when the conical angle in the TCS phase
above $H_{1}$ becomes rather small, at the field $H_{2}$ not that far from the
saturation field, further canting gains too little energy and it becomes
energetically favorable to flop back into the $ab$-plane, gaining back some of
the anisotropy energy. The resulting phase, found by minimization of the
minimal Hamiltonian, is a very unusual commensurate fan-like (FL) phase,
depicted in Fig. \ref{Fig4}(b). It can be described as a quadrupled structure
along the $c$-axis, with spins deviating from the $x$-direction, the direction
of the magnetic field, by the angles $\gamma,\gamma,-\delta,\delta
,-\gamma,-\gamma,\delta,-\delta$ which gradually decrease until the
forced ferromagnetic (FF) state, $\gamma=\delta=0,$ is reached (see \ref{SI3} for details). The FL phase
has a different periodicity for $M_{x},$ the projection of Mn moments onto the
$a$-axis, and for $M_{y}$, the projection onto the perpendicular axis. The
latter corresponds to $k_{c}$ = 0.25, the former to $k_{c}$ = 0.5, and the
variation of amplitude of $M_{x}$ is much smaller. The calculations [Fig.
\ref{Fig4}(c)] capture all features of the measured magnetization [Fig.
\ref{Fig2}(a)]. The predictions are also confirmed by our neutron data: the
first spin-flop from a nearly-helical to a nearly-cycloidal spiral leads to
about 50\% loss in the scattering intensity for (0, 0, $L$ $\pm$ $k_{n,z}%
$)-type Bragg peaks (neutrons do not scatter off the $M_{z}$ component in our
geometry when the scattering vector is along $L$), consistent with the discontinuous loss of intensity in the
experiment (see \ref{SI1}). In the minimal model the first spin-flop does not alter
the periodicity; experimentally, however, $k_{z}$ slightly increases in the TC
phase. To understand this, we need to step back to the reduced model that
retains separation of $K$ and $J^{z}.$ Then the MFT theory predicts a tiny
shortening of the spiral pitch at $H=H_{1},$ on the scale of $\approx
0.36J^{z}/J_{1}\sim1\%~$(see \ref{SI3}). The FL phase also finds full confirmation
in the experiment: at $H$ = $H_{2},$ as predicted, $k_{z}$ changes
discontinuously to $k_{c}$ = 0.25 and the predicted weaker satellite at
$2k_{c}$ is observed as well.

We now focus on the topological Hall effect (THE) and show its origin in a fluctuation-driven nematic chirality. The THE
appears in the TCS phase only. The fact that the THE is observed only at elevated
temperatures while the TCS phase exists in the entire temperature range below
330 K, strongly suggests a key role of thermal fluctuations. It is worth
remembering that the system is strongly 2D, with nearly two orders of
magnitude difference between the $ab$ and $c$ couplings. In this case, by
virtue of the Mermin-Wagner theorem, the mean field transition temperature of
several thousand K is dramatically suppressed by large and relatively slow
\textit{in-plane} fluctuations. This is reminiscent of the famous nematic
transition in the planar $J_{1}-J_{2}$ Heisenberg mode \cite{Chandra1990},
where these fluctuations can conspire in such a way to create a new,
non-magnetic order parameter without a long-range magnetic order. This
so-called nematic phase is realized in many Fe-based superconductors
\cite{Fernandes2014} and may be in other materials as well \cite{Zhang2017b}.

We will argue now that similar physics may be realized in the TCS phase. The
detailed theory is provided in \ref{SI4}. Here we present a summary of the results.
In a continuous approximation the TC spiral can be described as $\mathbf{M=M}%
_{x}\mathbf{+m}$, where $\mathbf{M}_{x}\Vert\mathbf{\hat{x}}$ is the induced
magnetic moment and is a constant, and $\mathbf{m\perp\hat{x}}$ is a cycloidal
spiral. It is also assumed that while the direction of the Mn moment can
change and fluctuate, the amplitude stays the same. The topological chiral
field given by the standard expression\cite{Nagaosa2013} $b_{x}~=~\mathbf{M}%
\cdot(\partial_{y}\mathbf{M}\times~\partial_{z}\mathbf{M})=\partial
_{y}\mathbf{M}\cdot(\partial_{z}\mathbf{M\times M})$ is thus zero in the TCS
phase (or in any phase) where $\partial_{y}\mathbf{M=0,}$ and hence there is
no THE. However, addition of a magnon fluctuation, propagating along $y$ with
wave vector $k_{y}$ gives $\mathbf{M~=~}M_{x}\mathbf{x+m+\mu}$ where the
fluctuating moments rotate in a plane defined by a vector $\mathbf{\omega,}$
such that $\omega\propto k_{y}.$ Then, $\frac{\partial\mathbf{M}}{\partial
z}~=~\frac{\partial\mathbf{m}}{\partial z}\mathbf{~=~}\mathbf{m}%
\times\mathbf{\hat{x}}$, and $\frac{\partial\mathbf{M}}{\partial y}%
~=~\frac{\partial\mathbf{\mu}}{\partial y}\mathbf{=\mu}\times\mathbf{\omega}$.
Using these equations on $b_{x},$ keeping only the terms quadratic in $\mu$,
and averaging over $y$ gives $b_{x}=-k_{y}m_{z}\mu^{2}$, and, unless
$\mathbf{\omega}||\mathbf{\hat{x},}$ $b_{x}\neq0.$

The physical meaning of this result is very simple; the TCS is one independent
magnon short of a chiral combination of static magnons. Of all possible
magnons there are some that generate positive chirality, but, by
crystallographic symmetry, for each such magnon there is a partner with the
same energy and opposite chirality. These two partners will be thermally
excited with the same probability and will cancel each other, in the absence
of an external field. However, they create non-zero chiral susceptibility,
reminiscent of the nematic susceptibility in Fe-based superconductors. We find
the chiral field in such a case to be:
\begin{equation}
\left\langle b_{x}\right\rangle =const\cdot TM_{z}^{2}H_{x}=const\cdot
(1-M^{2}/M_{s}^{2})TH_{x}.\label{R}%
\end{equation}
The topological Hall resistivity ($\rho^{T}$) is proportional to $\left\langle
b_{x}\right\rangle $, and hence $\rho^{T}$ can be calculated by Eqn. \ref{R}
using experimental parameters. It is to be noted that this expression is valid
only for $H_{1}<H_{x}<H_{2}$ and the topological Hall resistivity is zero
outside these limits. The theoretical $\rho^{T}$ is plotted together with the
measured data at $T$ = 245 K in Fig. \ref{Fig4}(d). The inset shows the
temperature dependence of $\rho^{T}$ at a constant field of 4 T, which is
linear in temperature as expected from Eqn. \ref{R} ($M_{z}$ depends on the
temperature very weakly, as one can see from the experimental data in Fig. \ref{SFH3}). The details of experimental and theoretical $\rho^{T}$
are provided in the \ref{SI5}. The remarkable agreement of the experimental data
with this phenomenological model provides insight into the microscopic origin
of the THE as stabilized by the thermal fluctuations creating an imbalance in
the right and left handed transverse conical spirals - a nematic spin
chirality. We want to point out that the exceptional agreement between theory and
experimental THE, may be to some extent fortuitous, given the simplicity of
the model and partitioning of the total $\rho_{H}$ (discussed in detail in
\ref{SI5}), but provides a strong support to the presented physical picture,
describing  the observed THE in terms of nematic chirality.

In summary, we have identified two unique magnetic phases, TCS and FL, in
YMn$_{6}$Sn$_{6}$, which emerge from the competitions between exchange
interactions, the magnetic anisotropies, and Zeeman energy, with a remarkable
agreement between bulk measurements, neutron diffraction, and first principles
calculations. The THE in the TCS phase is of particular interest. As opposed
to non-coplanar, and skyrmionic materials, this spiral magnet without static spin
chirality forms a non-zero internal skyrmionic magnetic field dynamically,
through preferential excitation of chiral fluctuations with a given
handedness. This field deflects the conducting charge and thus produces the
extra component to the Hall effect, the THE. We call this effect
``nematic chirality" by analogy with the nematic
phase in Fe-based superconductors. Our results not only provide a new THE
mechanism but also open promising avenues in looking for the chiral spin
textures in new materials and at temperatures relevant for practical applications.

\section*{Methods}

{\textbf{Crystal growth and characterization.}} Single crystals of YMn$_{6}%
$Sn$_{6}$ were grown by the self-flux method. Y pieces (Alfa Aesar 99.9 \%),
Mn pieces (Alfa Aesar 99.95 \%)  and Sn shots (Alfa Aesar 99.999 \%) were
loaded in a 2 mL aluminum oxide crucible in a molar ratio of 1:1:20. The
crucible was then sealed in a fused silica ampoule under vacuum. The sealed
ampoule was heated to 1175 $^{\circ}$C over 10 hours, homogenized at 1175
$^{\circ}$C for 12 hours, and then cooled to 600 $^{\circ}$C over 100 hours.
Once the furnace reached 600 $^{\circ}$C, the excess flux was decanted from
the crystals using a centrifuge. Well-faceted hexagonal crystals as large as
100 mg were obtained. The crystal structure of the compound was verified by
Rietveld refinement \cite{Mccusker1999} of a powder x-ray diffraction pattern
collected on pulverized single crystals at room temperature using a Rigaku
Miniflex diffractometer. The Rietveld refinement was carried out using
FULLPROF software \cite{Rodriguez-carvajal1993}.

{\textbf{Magnetic and transport property measurements.}} dc susceptibility
measurements were made using a Quantum Design VSM SQUID. dc magnetization and
transport measurements were measured using a Physical Property Measurement
System (PPMS). ac susceptibility measurements were carried out using a Quantum
Design Dynacool PPMS. Resistivity and Hall measurements were performed
following the conventional 4-probe method. Pt wires of 25 $\mu$m diameter were
attached to the sample with Epotek H20E silver epoxy. An electric current of 1
mA was used for the transport measurements. In magnetoresistance measurements,
the contact misalignment was corrected by field symmetrizing the measured data.

{\textbf{Neutron diffraction measurements.}} A single-crystal was oriented in
either the ($H, 0, L$) or ($H, H, L$) scattering plane on the triple-axis
neutron spectrometer, BT-7 \cite{Lynn2012} at the NIST Center for Neutron
Research (NCNR). Elastic diffraction data were taken with $E_{i}$ = $E_{f}$ =
14.7 meV and $25^{\prime}- 10^{\prime}- 10^{\prime}- 25^{\prime}$ full-
width-at-half-maximum (FWHM) collimators were used before and after the
sample, before the analyzer, and before the detector, respectively (unless
otherwise noted). A superconducting 7 T vertical field magnet system with a
top loading closed cycle refrigerator was used at the sample position such
that the applied field was parallel to the [1, $\bar{1}$, 0] crystallographic
direction. Bragg peaks were resolution limited and Gaussian in shape. Peaks
were therefore fit to Gaussians with the FWHMs constrained to be that of the
spectrometer resolution as determined by the program, ResLib
\cite{Zheludev2009}. Data using a superconducting 15 T vertical field magnet
system were taken in the ($H, H, L$) scattering plane, where the magnetic
field was also parallel to [1, $\bar{1}$, 0]. Moderately coarse resolution was
used with $-50^{\prime}-40^{\prime}R-120^{\prime}$ collimators (where
`R' indicates radial) and a position sensitive detector. Throughout the
manuscript, momentum is reported in reciprocal lattice units (r.l.u.) denoted
by using $H$, $K$, and $L$, where \textbf{Q} [$\mathring{A}$$^{-1}$] =
($\frac{4\pi}{\sqrt{3}a}H$, $\frac{4\pi}{\sqrt{3}a}K$, $\frac{2\pi}{c}L$).

{\textbf{First principles calculations.}} Most calculations were performed
using the projected augmented wave pseudopotential code VASP \cite{Kresse1996}%
, and the gradient-dependent density functional of Ref. \onlinecite{Perdew1996}. For
control purposes, some calculations were also repeated using the all-electron
linearized augmented plane wave code WIEN2k \cite{Blaha2001}. Hubbard
correlations were taking into account using the DFT+U with the fully-localized
double counting prescription, and the spherically averaged correction $U-J,$
with the value of $U-J$ given in the text.

\section*{Acknowledgements}

NJG acknowledges startup fund from George Mason University. Work in the
Materials Science Division at Argonne National Laboratory (JFM and JSJ) was
supported by the U.S. Department of Energy, Office of Science, Basic Energy
Sciences, Materials Science and Engineering Division. Work at ORNL (MAM) was
supported by the US Department of Energy, Office of Science, Basic Energy
Sciences, Materials Sciences and Engineering Division. The identification of
any commercial product or trade name does not imply endorsement or
recommendation by the National Institute of Standards and Technology. The authors thank Predrag Nikolic and Christian Batista for insightful 
discussions.

\section*{Author contributions}

NJG conceived and coordinated the project. NJG and NTM grew the crystals. NJG, NTM, DM and DCJ
characterized the samples. NJG and MAM performed the magnetic and
magnetotransport measurements. JFM contributed to the magnetic and transport
measurements. JSJ contributed to the transport measurements. LP, RLD and JWL
carried out neutron diffraction experiments. LP, MB, NJG and JWL performed
SANS experiment. IIM carried out the first principles calculations and devised
the phenomenological theory. NJG wrote the manuscript with contributions from
RLD, JWL and IIM. All authors contributed to the discussion of the results.

\begin{figure}[H]
\begin{center}
\includegraphics[scale=1]{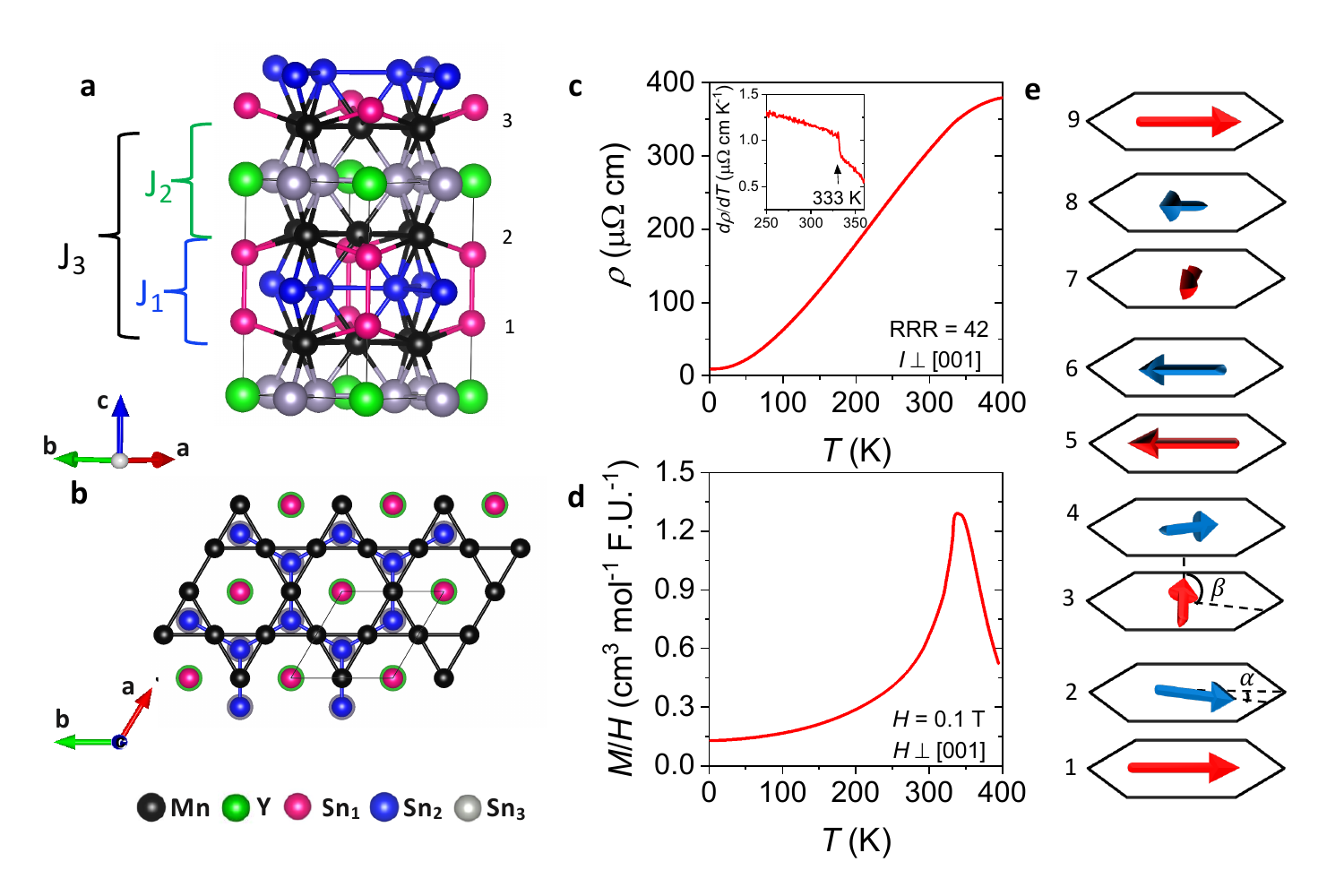}
\end{center}
\caption{\textbf{Crystal structure, and electrical and magnetic properties of
YMn$_{6}$Sn$_{6}$.}  a) Sketch of the crystal structure of YMn$_{6}$Sn$_{6}$.
b) Top view of the structure shown in panel (a). Within a unit cell shown by
the dark solid lines, there are two kagome planes with the formula Mn$_{3}$Sn
that are separated by Sn$_{3}$ and YSn$_{2}$ layers. J$_{i}$s are the exchange
constants between different Mn layers. c) Electrical resistivity of YMn$_{6}%
$Sn$_{6}$ as a function of temperature with the electric current applied in
the $ab$-plane. Inset shows the temperature derivative of the electrical
resistivity in the vicinity of T$_{N}$, which shows a jump at 333 K below
which an incommensurate spiral state develops. The residual resistivity ratio
(RRR = $\rho_{400 K}/\rho_{2 K}$) is 42 indicating a good sample quality. d)
Magnetic susceptibility ($M/H$) of YMn$_{6}$Sn$_{6}$ as a function of
temperature. e) Incommensurate magnetic structure of YMn$_{6}$Sn$_{6}$ in
the absence of external magnetic field. Arrows represent the direction of
ferromagnetic spins within a kagome plane. There is a small constant angle
$\alpha$ between the FM-coupled spins across the Sn$_{3}$ layer, and $\beta$
between the AF ones across the Sn$_{2}$Y which result in a spiral spin
arrangement, where every other Mn layer forms a spiral with the pitch defined
by $\alpha+ \beta\approx90{{}^{\circ}}$ and the two spirals rotated by
$\alpha$ with respect to each other. The incommensurate spirals repeat after
about four crystallographic unit cells or about nine Mn layers that are
indicated by the numbers 1 through 9.}%
\label{Fig1}%
\end{figure}

\begin{figure}[H]
\begin{center}
\includegraphics[scale=1]{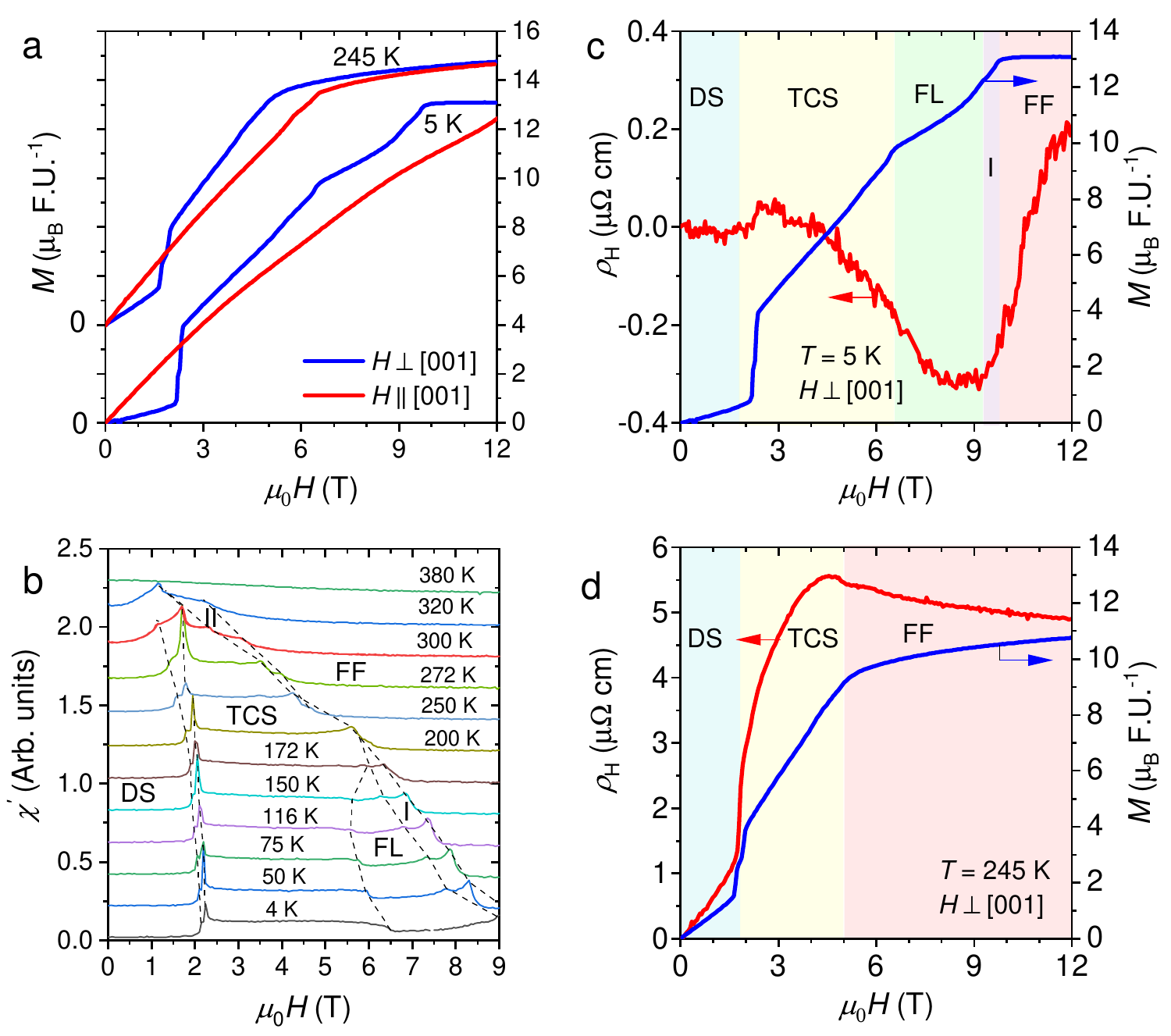}
\end{center}
\caption{\textbf{Magnetization and Hall effect of YMn$_{6}$Sn$_{6}$.} a)
Magnetization as a function of external magnetic field at 5 and 245 K with the
magnetic field applied parallel and perpendicular to the $c$-axis. Data have
been offset as indicated. b) Phase diagram of YMn$_{6}$Sn$_{6}$ constructed
from ac-susceptibility measurements. c) Hall resistivity (left axis) and
magnetization (right axis) as a function of magnetic field applied in the
$ab$-plane at 5 K, and d) at 245 K. DS, TCS, FL and FF stand for distorted spiral, transverse conical spiral,
fan-like, and forced ferromagnetic phases, respectively.}%
\label{Fig2}%
\end{figure}

\begin{figure}[H]
\begin{center}
\includegraphics[scale=.2]{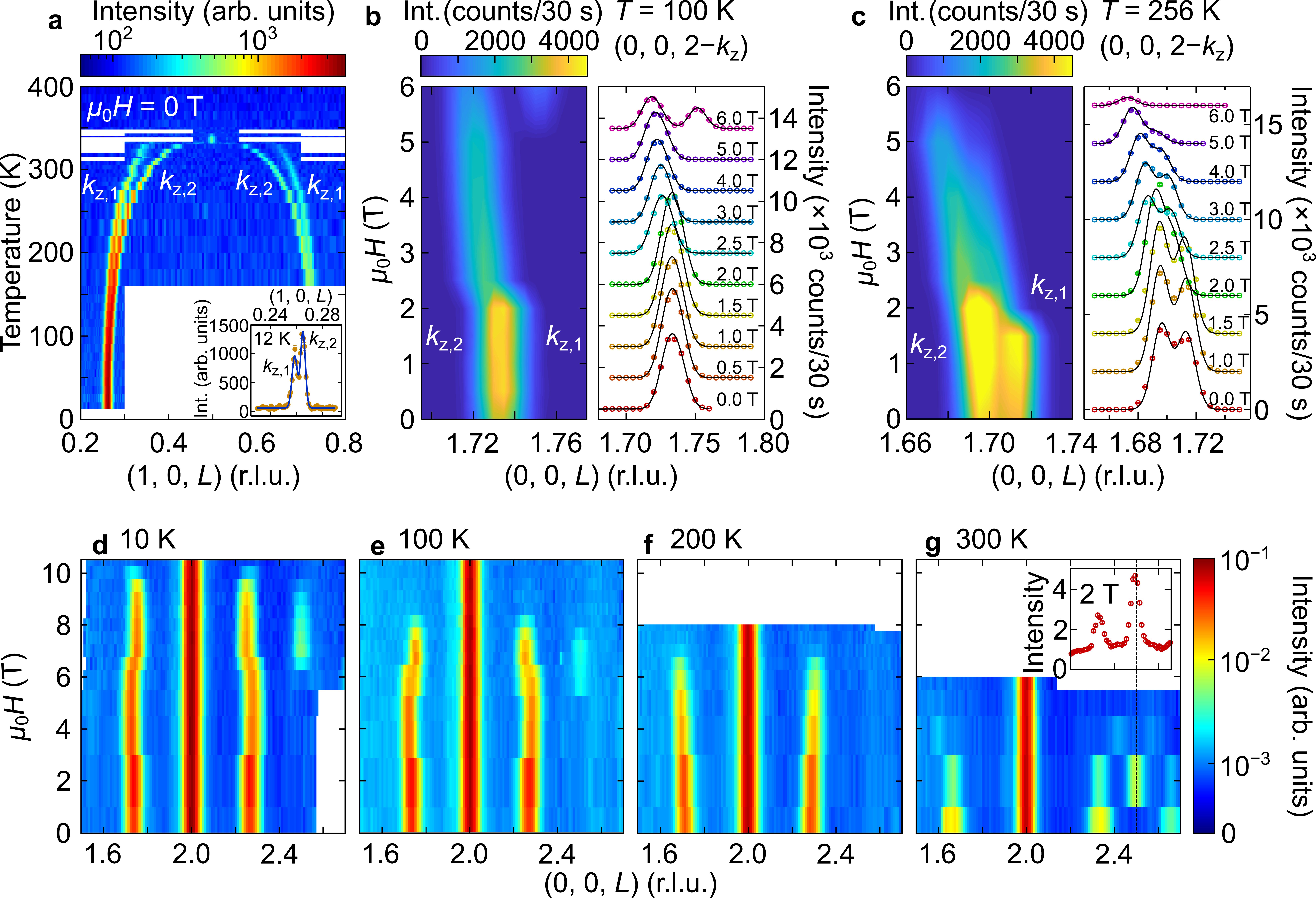}
\end{center}
\caption{\textbf{Single crystal neutron diffraction of YMn$_{6}$Sn$_{6}$.} (a)
Magnetic Bragg peaks tracked as a function of temperature. A commensurate
magnetic peak at $L = 0.5$ appears between 345 K and 330 K, and the two
incommensurate magnetic structures stemming from the wave vectors $k_{z,1}$ and $k_{z,2}$
appear at 330 K and persist to the base temperature measured, 12 K. The inset,
taken with high instrumental resolution, shows that the two wave vectors do
not converge, even as they get closer with decreasing temperature. (b)-(c)
Incommensurate magnetic Bragg peaks (0, 0, $2-k_{z,n}$) ($n=1,2$) tracked at 100 K
and 256 K, respectively, as a function of applied magnetic field. The solid
black lines in the right-hand panels of (b) and (c) are Gaussian fits to
the data described in Methods. An offset was added between individual $L$
scans for clarity. Offsets are 1500 counts/30 sec. for (b) and 2000 counts/30
sec. for (c). (d)-(g) Neutron diffraction data taken up to higher fields with
a position sensitive detector and coarse resolution for (d) 10 K, (e) 100 K,
(f) 200 K, and (g) 300 K. In these data, $k_{z,1}$ and $k_{z,2}$ are not
resolvable, but the high fields at which the data were taken reveal the field
ranges at which each of the magnetic phases are present. The inset of (g) is a
cut taken from the main panel at 2 T, where the dashed black line shows that
the new peak appearing at this field is commensurate at $L = 2.5$.}%
\label{Fig3}%
\end{figure}

\begin{figure}[H]
\begin{center}
\includegraphics[scale=1]{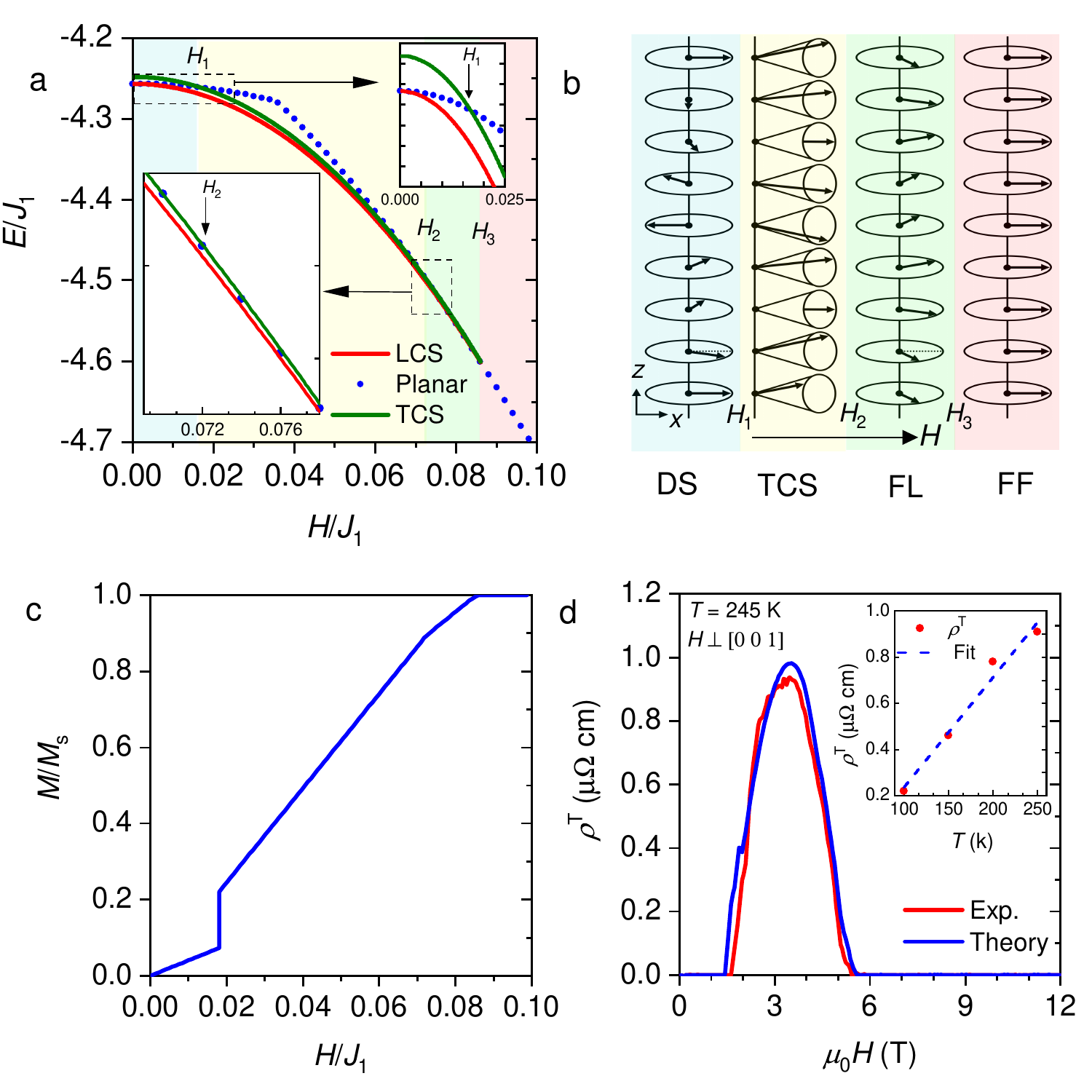}
\end{center}
\caption{\textbf{First principles calculation and phenomenological model of
nematic chirality for topological Hall effect.} a) Energy for different
magnetic states as a function of reduced magnetic field obtained in MFT
calculations. The planar state represented by the blue dotted line, has DS structure below the kink and FL structure above the kink.  b) Sketch of different field-induced magnetic structures. DS, TCS, FL and
FF stand for distorted spiral, transverse conical spiral,
fan-like and forced ferromagnetic phases. c) Calculated magnetization as a
function of in-plane field. d) Experimental and theoretical topological Hall
resistivity as a function of external magnetic field at 245 K. Inset shows the
temperature variation of the THE at 4 T below 250 K. The dashed line is a
linear fit to the experimental data.}%
\label{Fig4}%
\end{figure}

\widetext
\begin{center}
\pagebreak
\hspace{0pt}
\vfill
\textbf{\large Supplementary Information for: \\Novel magnetic states and nematic spin chirality in the kagome lattice metal
YMn$_{6}$Sn$_{6}$}
\\[12 pt]
Nirmal J. Ghimire$^{\ast}$, Rebecca L. Dally, L. Poudel, D. C. Jones, D. Michel, N. Thapa Magar, M. Bleuel, Michael A. McGuire, J. S. Jiang, John F. Mitchell, Jeffrey W. Lynn, and I. I. Mazin
\vfill
\hspace{0pt}
\pagebreak
\end{center}


\setcounter{equation}{0}
\setcounter{figure}{0}
\setcounter{table}{0}
\setcounter{page}{1}
\makeatletter

\renewcommand\thesection{SI\arabic{section}}
\renewcommand{\theequation}{S\arabic{equation}}

\renewcommand{\thetable}{S\arabic{table}}
\renewcommand\thefigure{S\arabic{figure}}
\renewcommand{\theHtable}{S\thetable}
\renewcommand{\theHfigure}{S\thefigure}

\section{Neutron diffraction}\label{SI1}
One of the central motivations for investigating the properties of YMn$_6$Sn$_6$ was to determine the origin of the observed topological Hall Effect (THE).  One possibility was that a skyrmion lattice formed in the system, and therefore our first neutron experiments were using small angle neutron scattering (SANS) to search for skyrmions with the NG-7 SANS instrument at the NCNR using a 9 T horizontal field magnet ($H~||~[ 1, 1, 0]$) and exploring the temperature range from 4 K to 300 K.  Figure \ref{SFN6}(a) shows a background subtracted image over a wide wave vector range.  No large scale magnetic structures were observed such as expected for a skyrmion lattice.  Rather, just a single diffraction spot ­­­at a rather large (for SANS) wave vector was observed, indicating that long range magnetic order with an incommensurate modulation is realized.  Fig. \ref{SFN6}(b) shows an example of the field dependence of the magnetic order, demonstrating a complex series of magnetic phase transitions.  Figure \ref{SFN6}(c) shows cuts through the scattering at a field of 7 T for a series of temperatures, where it was first observed that there are two closely-spaced incommensurate wave vectors.  To further elucidate the nature of the magnetic structures as a function of both magnetic field and temperature, wide angle diffraction data were collected as we now describe.

Figure \ref{SFN1} shows the variation of the incommensurate
Bragg peaks at (1, 1, $0-k_{z,n}$) with the external magnetic field applied
in the $ab$-plane at 100 K and 256 K, similar to the data presented about (0, 0, $2
-k_{z,n}$) in Figs. 2(b) and (c). The momentum resolution in the
measurement about the (1, 1, $0-k_{z,n}$) peaks is much better than that about the (0, 0,
$2-k_{z,n}$) peaks, and as a result, the separation of $k_{z,1}$ and $k_{z,2}$ at 100
K is clearer in the former case. Determination of peak centering obtained from
the fitting of (1, 1, 0 -$k_{z,n}$) is thus unambiguous. We used the distance
between the peak centers determined from the fitting of (1, 1, $0-k_{z,n}$) to
distinguish the behavior of the two peaks in (0, 0, $2-k_{z,n}$) discussed in
the main text. Additionally, the FWHM of the Gaussian fits were fixed to be
that of the instrumental resolution as described in the Methods section. The
integrated intensity obtained from the Gaussian fits are shown in
Figs. \ref{SFN2}(b) and (c). It can be seen from these figures
that the percent decrease in intensity at the metamagnetic transition $H_1$ is not as great as the (0, 0, $2-k_{z,n}$) peaks, as expected for a helical to cycloidal transition. Intensities of the peaks measured at (1, 1, $0-k_{z,n}$) are much
weaker compared to the peaks measured at (0, 0, $2-k_{z,n}$), but it is still clear that at
6 T, $k_{z,2}$ has practically no intensity and the intensity of $k_{z,1}$ is
very weak. As the commensurate peaks emerge at 6 T [Figs. \ref{Fig3}(b) and \ref{SFN2}(a)], it suggests that they
appear at the cost of the incommensurate peaks.

In Fig. \ref{SFN3}, we plot the experimental ratio of intensity
above versus below $H_{1}$ together with the corresponding calculated ratios.
This plot supports the first principles calculations results discussed in the
main text of helical to cycloidal spin flop at $H_{1}$. Given a $k_{z}$ and
any $\alpha$ value (the rotation angle between Mn layers separated by pure Sn
layers), the calculated cycloidal:helical intensity ratio is 0.5 for (0, 0, $L$)-type
peaks. This is because in the helical state, all spins lie perpendicular to
(0, 0, $L$)-type scattering vectors, and neutrons are only sensitive to the
component of a spin which is perpendicular to the scattering vector. When the
moments flop into the ($1, 1, 1$) plane and the structure becomes cycloidal,
half of the total spin magnitude is now projected along (0, 0, $L$), thus
reducing the intensity by half. There is generally much less of an intensity
suppression after the transition for Bragg peaks at ($H, H, L$) positions
where $H$ is not equal to 0. This is because for a specific $k_{z}$, the
cycloidal:helical structure factor ratios for a given peak do not change regardless of the value of $\alpha$. There are, however,
slight differences in the ratios for different wave vectors, although they are
small for the range of $k_{z}$ observed in YMn$_{6}$Sn$_{6}$. The calculations
used the different $k_{z,n}$ values from the neutron data for the different
temperatures and on either side of $H_{1}$. The (0, 0, $L$)-type peaks
decrease in intensity more relative to the ($H, H, L$) ($H \neq0$) peaks
across the $H_{1}$ boundary, consistent with what would be observed for a
helical to cycloidal spin-flop transition. Combined with the theoretical
results, the neutron data do support the proposed helical to cycloidal
spin-flop transition at $H_{1}$, especially when considering other complex
incommensurate magnetic structures, such as the fan or spin density wave,
which would bear the hallmark of Bragg peak harmonics and any harmonics are
absent in the data presented here.

Additionally, we expect that immediately following the spin-flop transition,
the cycloidal arrangement of spins would begin to cant along \textbf{H}, forming a
transverse conical structure. We confirmed this canting by tracking the net
induced ferromagnetic component at the (0, 0, 4) Bragg peak as a function of applied magnetic
field. In zero field, the intensity is purely nuclear in origin, with a
field-induced intensity simply adding to the structural part. We see an abrupt
emergence of an induced net ferromagnetic component at $H_{1}$ for the 100 K and 256
K data, as shown in  Figs. \ref{SFN4}(a), and (b), respectively. We note that
the 256 K data in Fig. \ref{SFN4}(b) were taken with coarser instrumental
resolution as can be discerned by the $x$-axes in Figs. \ref{SFN4}(a) and (b).
The integrated intensity with respect to the zero-field intensity is shown for
both temperatures in Fig. \ref{SFN4}(c). The abrupt increase in intensity is
indicative of a sudden canting of the moments towards the applied field
direction, $[1, \bar{1}, 0]$, and the magnitude of this projection, in $\mu_{B}^{2}%
$Mn$^{-1}$, is displayed on the right axis of Fig. \ref{SFN4}(c). Note that
the induced moment along the applied field direction is smaller at the higher
temperature as expected, and both behaviors are completely consistent with
bulk magnetization measurements presented in the main text [Fig. \ref{Fig2}(a)].
\section{First principles calculations}\label{SI2}
Total energies of 10 different collinear ($\alpha$ and $\beta$ defined below are
either 0 or 180$%
{{}^\circ}%
)$ magnetic patterns were calculated in a supercell containing 8 Mn layers.
Individual layers were ordered FM, and the selected patterns were: $uddddddu,$
$udddddud,$ $uddddudd,$ $uddduuud,$ $uddduuud,$ $ududdudu,$ $udududdu,$
$udududud,$ $uudduudd,$ $uuduuudd,$ $uuuuuuuu,$ where $u$ stands for an up-polarized layer, and $d$ for down-polarized. These were fitted to the
following Hamiltonian:
\begin{align}
2E  &  =J_{1}\cos\alpha+J_{2}\cos\beta+2J_{3}\cos(\alpha+\beta)\nonumber\\
&  +J_{4}\cos(2\alpha+\beta)+J_{5}\cos(\alpha+2\beta)+2J_{6}\cos
(2\alpha+2\beta) \label{H8}%
\end{align}
where $J_{i}$ are exchange interactions defined in Fig. \ref{FigJ}, $E$ is the energy per layer, and
the angles $\alpha$ and $\beta$ define rotations between the planes bridged by
Sn$_{3}$ or Sn$_{2}$Y layers [see Fig. \ref{Fig1}(a)]. At low temperature the wave vector is
$\approx\pi/2c.$ Fitting quality is very high, as shown in  Fig. \ref{fitquality} (blue circles). {The resulting values for $J_{1-6}$ are
}${-50.9,~9.1,~}4.1,~-4.0,~-7.5,$ and $-3.4$ meV{.} An immediate observation
is that, indeed, $J_{1}$ is strongly FM, $J_{2}$ and $J_{3}$ are both AF, but
(i) their absolute values are quite far from the stability range of any
spiral and (ii) longer-range interactions, especially $J_{5},$ are very
important. A closer look reveals that the main factor preventing the formation
of a spiral is the large $J_{1}.$ Indeed, for a FM $J_{1}$, a spiral can only
be stable if $J_{2}/2(|J_{1}|+J_{2})<J_{3}/J_{1}<J_{2}/2(|J_{1}|-J_{2}).$
This range, for $J_{2}\ll J_{1},$ becomes infinitely narrow,
$J_{2}-J_{2}^{2}/J_{1}<2J_{3}<J_{2}+J_{2}^{2}/J_{1}.$ Including the long-range
interactions does not change this picture: the ground state is the collinear
$uudd$ pattern.

In many systems, the addition of Coulomb correlations reduces short-range magnetic
couplings, but less so the long range ones. With this in mind, we repeated the
calculations by adding a Hubbard $U$ in the common LDA+U approximation. Figure
\ref{fitquality} (symbols other than blue circles) shows the fit quality,
which is still high. Figure \ref{UJ} shows how the fitted parameters vary as a
function of the effective $\tilde{U}=U-J.$

It is instructive to look at the phase diagram in the \textquotedblleft
reduced\textquotedblright\ ($J_{1},$ $J_{2},$ $J_{3}$ only) model. Figure
\ref{GS} shows where our calculated effective exchange parameters fall at
different $\tilde{U}\approx0.4$ eV and for $\approx0.6$ eV indeed a spiral is stable,
and another for very large (likely unphysical for a good metal) $U.$ It is
also useful to compare the calculated magnetic moments on Mn with the
experimental number of { $M_{\exp}=2.1$}. {The fact that even at $\tilde{U}=0$ the
moment is overestimated indicates a strongly fluctuating system and weak
correlations. }

On the other hand, {an effective magnetic moment of $\mu_{eff}$ = 3.6 $\mu_{B}$
was extracted from the high-temperature susceptibility\cite{Venturini1991s},
consistent with a spin $S=3/2$ (magnetic moment 3 $\mu_{B}).$ Thus one can
conclude that, despite the fact that for $\tilde{U}\sim2-3$ eV the system re-enters
a spiral region, this part of the calculated phase diagram is unphysical. However, the range of 0.4 -- 0.6 eV yields $M\sim2.7$ $\mu_{B}$ (Fig. \ref{M})
consistent with $\mu_{eff},$ and$,$ after being reduced by fluctuations,
with $M_{\exp}$.} As Fig. \ref{angles} illustrates, the calculated spiral
angles at $\tilde{U}=0.4$ eV are $\alpha=-22%
{{}^\circ}%
$ and $\beta=138%
{{}^\circ}%
.$ Aside from some overestimate in $\beta,$ the overall agreement is good.

Plotting $J_{i}$ as a function of distance clearly shows that it is
inconsistent with the RKKY expression; in particular, it decays much more
slowly than $1/d^{4}.$ Nevertheless, there is no question that the
interaction is transferred by conduction electrons. The conclusion is then
that the Fermi surface must be rather complicated, and indeed it is. Another
complication is that the standard RKKY formalism describes the interaction of
localized moments in a nonmagnetic matrix. This is, obviously, inapplicable
here. The closest analogy would be to start with a FM state and see whether it
may be unstable against the formation of a spin density wave (SDW). To address this, we computed the
Fermi surface, which is shown in  Figs. \ref{FS}(a) and (b).

Examining the Fermi surfaces, we see immediately that the spin-down surface is
somewhat 2D and does not bear any obvious signature of nesting. {The spin-up
surface, on the other hand, is rather 3D, and has two pockets, one electron
and one hole, which nest rather well with $q_{z} (\AA^{-1})=0.235\frac{2\pi}{c},$ which
agrees reasonably with the spiral vector in the experiment. }This is
illustrated in Fig. \ref{FS}(c), showing a 2D cut of the pockets in question.
Of course, this instability, which is similar in spirit, but rather different
in details from RKKY, is superimposed on top of other, short range
interactions, which affect the final outcome.

\section{Mean field theory in external magnetic field}\label{SI3}
Previous analyses were based on the assumption that only two ground states
compete, a longitudinal conical spiral (LCS), where the field is oriented along the
spiral vector, and the distorted spiral (DS)\cite{Rosenfeld2008as} where the field is applied normal to the spiral pitch. In the
former, each moment is rotated out of the plane by the same amount to form a
component parallel to the field. In the latter, the moments rotate in the
plane, but retain the general spiral structure. We will show that this does
not exhaust possible magnetic states.

Analyzing the complete model of Eq. \ref{H8} with an external field in an
arbitrary direction is too cumbersome; for
simplicity, we will reduce the model to the \textquotedblleft
standard\textquotedblright\ $J_{1}-J_{2}-J_{3}$ one, keeping in mind that the
physics of the spin-flop and spin-flip transitions is roughly the same. We
then adjust the parameters to generate a SDW with $q\approx(0,0,0.25),$ and the
pitching angles $\alpha=-20%
{{}^\circ}%
$ and $\beta=110%
{{}^\circ}%
$ ($\alpha+$ $\beta=90%
{{}^\circ}%
),$ that is, $J_{2}/J_{1}=-$0.364, $J_{3}/J_{1}=0.171$ (and, $J_{1}<0).$
For reference, the angles are in agreement with published analyses\cite{Rosenfeld2008as},%
\begin{align}
\alpha &  =-\mathrm{sign}(J_{1}J_{3})\cos^{-1}\left(  \frac{J_{2}J_{3}}%
{J_{1}^{2}}-\frac{J_{3}}{J_{2}}-\frac{J_{2}}{4J_{3}}\right)  ,\\
\beta &  =\cos^{-1}\left(  \frac{J_{3}J_{1}}{J_{2}^{2}}-\frac{J_{1}}{4J_{3}%
}-\frac{J_{3}}{J_{1}}\right)  ,\label{ab}\\
\alpha+\beta &  =\cos^{-1}\left(  \frac{J_{1}J_{2}}{8J_{3}^{2}}-\frac{J_{2}%
}{2J_{1}}-\frac{J_{1}}{2J_{1}}\right).
\end{align}

First, let us calculate the energy of the LCS state. Taking the same rotation
angle $\theta$ for all moments and assuming the ideal in-plane FM order, we
can write the total energy per one 1$\times$1$\times$4 supercell as
\begin{equation}
E_{LCS}=4(J_{1}\cos\tau-J_{2}\sin\tau)M_{||}^{2}+4(J_{1}+J_{2}+2J_{3})M_{\perp
}^{2}-8HM_{\perp} \label{US}%
\end{equation}
where $\tau=-\alpha=\beta-90%
{{}^\circ}%
=20%
{{}^\circ}%
,$ $M_{||}=\cos\theta,$ and $M_{\perp}=\sin\theta$ (as before, we normalize
all interactions to unit moment). In the following we shall simplify notations
by using $e=E/|J_{1}|,$ $j_{2,3}=J_{2,3}/J_{1},$ and $h=H/|J_{1}|.$ The sign is
kept in the second definition to harmonize notations with Refs. \onlinecite{Rosenfeld2008as} and \onlinecite{Rozenfeld2009s}. Minimizing with respect to $\theta,$ and using the selected
parameters, we find
\begin{equation}
e_{LCS}=-4.257-46.443h^{2}.
\end{equation}
The angle $\theta$ changes gradually from $\pi/2$ to 0, and saturates at $h=$0.086.
Next, we consider the field applied in the plane. Since the leading wave vector at low temperature, experimentally is close to (0, 0, 0.25), and changes little with magnetic field (notwithstanding important, but small changes), we will consider a $commensurate$ SDW in the $1\times1\times4$ supercell, and, contrary to the
previous works, we shall assign different angles $\phi_{i}$ to each of the
eight sites. In the absence of a field, $\{\phi_{2},\phi_{3},\phi_{4},\phi
_{5},\phi_{6},\phi_{7},\phi_{8}\}-\phi_{1}=\{-\tau,\pi/2,\pi/2-\tau,\pi
,\pi-\tau,3\pi/2,3\pi/2-\tau\}.$ The total energy now looks like ($\phi_{9}=\phi_{1})$
\begin{align}
e_{DS} &  =-\sum_{j=1,4}\cos(\phi_{2j}-\phi_{2j-1})-j_{2}\sum_{j=1,4}\cos
(\phi_{2j}-\phi_{2j+1}) \nonumber\\
&  -j_{3}\sum_{i=1,8}\cos(\phi_{i}-\phi_{j+2})-h\sum_{i=1,8}\cos(\phi_{i}).
\end{align}
Minimizing this expression with respect to $\phi_{i}$ reveals two interesting
transition [Figs. \ref{Fig4}(a) and (c)]. At $h=h_{f}\approx0.035$ the
energy slope changes discontinuously, $i.e.,$ the magnetization experiences a
jump. At $h=h_{3}=$ 0.086  (which represents $H_3$ in the phase diagram), the moment saturates, and therefore $\chi=dM/dH$
has a discontinuity. A closer inspection reveals that up to $h_{f}$ the
grounds state is indeed a slightly distorted spiral (DS). Between $h_{f}$ and
$h_{3},$ however, it is a qualitatively different state, which we call
fan-like (FL) phase, where the moments 1 and 2 are aligned
ferromagnetically, and so are 5 and 6. They are gradually rotating with the
field until they become parallel to the latter at $h_{3}.$ At the same time,
the pairs 3 and 4, and 7 and 8 are canted from the field in opposite
directions. {Thus, the moments form the following angles with the field:
($\gamma,\gamma,-\delta,\delta,-\gamma,-\gamma,\delta,-\delta).$ Immediately
after the transition their values are $\gamma=77.25%
{{}^\circ}%
$ and $\delta=-10.06%
{{}^\circ}%
.$ The normalized energy $e=E/J{_1}$ as a function of the normalized field $h=H/J_{1}$ of the LCS and planar (the planar state is DS below, and a FL structure above the kink at $h_f$) phases are shown by the solid red line and dotted blue line, respectively in Fig. \ref{Fig4}(a) of the main text.

{However, without taking into account any anisotropy, $E_{LCS}$ is alway lower
than either $E_{DS}$ or }$E_{FL},${ so the former would immediately flop and
stay as such at all fields.}
So, let us include an easy-plane anisotropy, by adding a penalty term
$KM_{\perp}^{2}$. The DS state is not affected. The spin-flopped LCS, competing with DS, is the transverse conical spiral
(TCS), as described in the main text. Its energy is
\begin{align}
E_{TCS} &  =4(J_{1}\cos\tau-J_{2}\sin\tau)M_{||}^{2}+4(J_{1}+J_{2}%
+2J_{3})M_{\perp}^{2}\nonumber\\
&  -8HM_{\perp}+8KM_{\perp}^{2}.%
\end{align}
Since the average value of $M_{\perp}^{2}$ in the TCS state is $M^{2}/2,$ the
penalty term for $h=0$ is 8$KM^{2}/2=4K^{2}$ (normalizing to $M=1).$ This
penalty will gradually decrease with $h,$ as the canting toward the field
direction increases. In short, it amounts to just adding a penalty term equal
to $const\times\lbrack1-(h/h_{sat})^{2}]$ to Eq. \ref{US}. Energy of the TCS phase as a function of the magnetic field will then be different, which is shown in Fig. \ref{Fig4}(a) as the green solid line.

At large fields, close to saturation, the energy gain derived from
a larger spin susceptibility in the TCS phase is nearly lost, and at some
critical field $h_{2} \approx 0.072$ ($H_2$ in the phase diagram) it becomes energetically favorable to regain the
anisotropy energy by flopping again into the $ab$-plane, into the FL phase (of
course, the angles $\gamma$ and $\delta$ are now very small). 

So the magnetic phases obtained are:\\
\noindent
1) For the magnetic field along the $c$-axis: only the LCS phase is possible, and it gradually
changes until the saturation is reached at $h=h_{3}.$ 2) For the magnetic field in the $ab$-plane: {At very small fields the state is $DS.$}
In this phase, magnetization increases with a rather small slope until the spin-flop field,
proportional to $\sqrt{K},$ is reached ($K=0.01J_{1}$ was used in the plot,
inspired by the calculated value of $K\sim0.2$ meV, and then $h_{flop}%
=h_{1}\approx0.018),$ at which point the state discontinuously transforms
into the TCS phase $via$ a spin flop. Highly unusual, it flops again at the
field $h_{2}$ into the FL phase (again, for our selection of $K,$ it
is $\approx0.072),$ and finally saturates at $h_{3}=0.086.$

Let us now estimate the effect of the two spin-flops at, $h_{1}$, and $h_{2}$, on the spiral vector. In the
above calculations we absorbed all magnetic anisotropies into one single-site term. However,
there are no \textit{a priori} arguments that anisotropic exchange (of the form
$J^{z}M_{i}^{z}M_{j}^{z})$ should be small compared to the single-site anisotropy.
On the contrary, cases are known, when light magnetic $3d$ ions are bridged by
heavy nonmagnetic elements (as Sn and Y in our case) and the anisotropic
exchange dominates, for instance in CrI$_{3}$, which, as well as YMn$_{6}$Sn$_{6},$ has a large magnetic moment.

Dividing the magnetic anisotropy into the single-site and exchange parts would not
change the calculated phase diagram, except for one aspect. Indeed, while the
onsite anisotropy does not change, in the lowest order, the spiral pitch, the
anisotropic exchange does. Essentially, it adds an $antiferromagnetic$
component to the ferromagnetic bonds and a $ferromagnetic$ component to the
antiferromagnetic bonds. Assuming that the leading contribution comes from the
largest $J$ (which is ferromagnetic $J_{1}),$ one can calculate the derivative
$d\cos(\alpha+\beta)/dJ_{1}=J_{2}/8J_{3}^{2}+J_{2}/2J_{1}^{2}-1/2J_{2}%
=J_{1}^{-1}(j_{2}/8j_{3}^{2}+j_{2}/2j_{1}^{2}-1/2j_{2}=-0.364J_{1}^{-1}.$
Adding an antiferromagnetic contribution to $J_{1}$ (i.e., $\Delta J_{1}>0)$
will have a negative effect $\cos(\alpha+\beta),$ that is to say,
$\alpha+\beta$ will be larger, and so will be the spiral $q,$ in agreement
with the experiment.

Calculations give $K\approx-0.12$ meV/Mn (easy axis) and $(J_{1}^{z}+J_{2}%
^{z})/2\approx0.34$ meV (easy plane). This contribution is positive, i.e.,
antiferromagnetic. Assigning it entirely to $J_{1},$ we get $J_{1}^{z}$ of the
right sign, and $J_{1}^{z}/J_{1}\sim0.34/51\sim0.7\%,$ in qualitative
agreement with the experiment.
\section{Theory of nematic chirality in YM\lowercase{n}$_{6}$S\lowercase{n}$_{6}$} \label{SI4}
We now present a phenomenological theory of fluctuation-generated chirality
and the topological Hall effect (THE) on a background of a static cycloidal or
transverse conical magnetic spiral, which is observed at finite temperatures
in an external magnetic field. We will assume that the amplitude of the Mn
moments is constant ($i.e.|\mathbf{M|}=1$). The four phases of interest in
YMn$_{6}$Sn$_{6}$ are [see Figs. \ref{Fig2} and \ref{Fig4}(a)]: \newline\noindent1. A
longitudinal conical spiral (LCS) propagating along $z$ in the field parallel
to $z,$ where the magnetic moment is described, in a continuous approximation,
as%
\begin{align}
\mathbf{M}_{~} & \mathbf{=M}_{z} +\mathbf{m}\\%
\mathbf{M}_{z}  &  =const\\%
\frac{\partial\mathbf{m}}{\partial z}  &  =\mathbf\mathbf{m}%
\times\mathbf{z,}%
\end{align}
where $\mathbf{M}_{z}$ $||$ $\mathbf{z,}$ and $\mathbf{m\perp z.}$ Here and
below $\mathbf{z}$ is the unit vector. \newline\noindent2. A distorted spiral
(DS) that appears in the in-plane field (parallel to $x,$) and below the first
spin-flop ($i.e.H_{1}$) transition. In the first approximation,%
\begin{align}
\mathbf{M}_{~}  & =\mathbf{\tilde{M}}+\mathbf{m}\\
\frac{\partial\mathbf{\tilde{M}}}{\partial z}  &  =\mathbf{\tilde
{M}}\times\mathbf{z}\\
\mathbf{m}_{~}  &  =\tilde{M}_{y}\mathbf{\tilde{M}}\times\mathbf{z,}%
\end{align}
where $\mathbf{\tilde{M}\perp z}$ and $\mathbf{m\perp z.}$ \newline\noindent3.
A fan-like (FL) phase. \newline\noindent4. A transverse conical spiral (TCS)
that appears in the in-plane field (parallel to $x:$)
\begin{align}
\mathbf{M}_{~}  &  =\mathbf{M}_{x}+\mathbf{m}\\
\mathbf{M}_{x}  &  =const\\
\frac{\partial\mathbf{M}}{\partial z}  &  =\frac{\partial\mathbf{m}}{\partial
z}\mathbf{=}\mathbf{m}\times\mathbf{x,}%
\end{align}
where $\mathbf{M}_{x}~||~\mathbf{x,}$ and $\mathbf{m\perp x.}$

We\ will now use the standard expression for the topological field, see
\textit{e.g.}, Eq. B4b in Ref. \onlinecite{Nagaosa2013s} (omitting the
coefficient of 2):%
\begin{equation}
b_{\alpha}~=\varepsilon^{\alpha\beta\gamma}~\mathbf{M}\cdot(\partial_{\beta
}\mathbf{M}~\times~\partial_{\gamma}\mathbf{M}).\label{N}%
\end{equation}

In the first case described above, that is, for an external field in the
$z$-direction, the field couples with $b_{z,}$ and%
\begin{equation}
b_{z}~=~\mathbf{M}\cdot(\partial_{x}\mathbf{M}~\times~\partial_{y}\mathbf{M}).
\end{equation}
In the ground state both $\partial_{x}\mathbf{M}~$and $\partial_{y}%
\mathbf{M}=0\mathbf{,}$ so generating a nonzero $b_{z}$ requires exciting
simultaneously two types of magnons, with two different in-plane vectors,
which is an unlikely case.

In cases 2-4, $i.e.,$ for an external field along $x$-direction, only $b_{x}$
couples with the external field, so without losing generality we can rewrite
it as:
\begin{equation}
b_{x}~=~\mathbf{M}\cdot(\partial_{y}\mathbf{M}\times~\partial_{z}%
\mathbf{M})=\partial_{y}\mathbf{M}\cdot(\partial_{z}\mathbf{M\times M}).
\end{equation}

In the ground state $\mathbf{M}$ only varies with $z,$ so this expression is
obviously zero. In a planar-helical ($\mathbf{H=0)}$ state $\partial
_{z}\mathbf{M\times M}$ is parallel to $z,$ and averages to zero over all
planes. So, let us concentrate on the state (4). Let us then write
$\mathbf{M}$ as%
\begin{equation}
\mathbf{M=}~M_{x}\mathbf{x+m+\mu}%
\end{equation}
where, as mentioned, $\mathbf{m\perp x,}$ and $\mathbf{\mu}$ represents a
magnon propagating along $y$ (in order to have nonzero $\partial
_{y}\mathbf{M)}$ in a plane defined by a vector $\mathbf{\omega,}$ such that
$\omega\propto k_{y}.$ Then%
\begin{align}
\frac{\partial\mathbf{M}}{\partial z}  &  =\frac{\partial\mathbf{m}}{\partial
z}\mathbf{~=}~\mathbf{m}\times\mathbf{x}\\
\frac{\partial\mathbf{M}}{\partial y}  &  =\frac{\partial\mathbf{\mu}%
}{\partial y}\mathbf{~=\mu}\times\mathbf{\omega}%
\end{align}
Let us now calculate $b_{x}$%
\begin{align}
b_{x}  &  =\mathbf{(}M_{x}\mathbf{x+m+\mu)}\cdot((\mathbf{\mu}\times
\mathbf{\omega})~\times~(\mathbf{m}\times\mathbf{x}))\\
&  =\mathbf{\mu}\cdot([(\mathbf{\mu}\times\mathbf{\omega})\cdot\mathbf{x]m-}%
[(\mathbf{\mu}\times\mathbf{\omega})\cdot\mathbf{m]x})
\end{align}
where the terms linear in $\mu$ are dropped because they average to zero upon
integrating over $y.$ Continuing with the expansion,%
\begin{equation}
b_{x}=(\mathbf{\mu}\cdot\mathbf{m)}[\mathbf{\mu\cdot(\omega}\times
\mathbf{x)]-}(\mathbf{\mu}\cdot\mathbf{x)}[\mathbf{\mu\cdot(\omega}%
\times\mathbf{m)]}%
\end{equation}
It is important to note that we can consider each $ab$ plane independently, as
they are capable of fluctuating independently. Let us for simplicity consider
the case $\mathbf{\omega~||~z.}$ Then%
\begin{equation}
b_{x}=(\mathbf{\mu}\cdot\mathbf{m)}[\mathbf{\mu\cdot y]-}(\mathbf{\mu}%
\cdot\mathbf{x)}[\mathbf{\mu\cdot(\omega}\times\mathbf{m)]=}k_{y}m_{z}\mu^{2}%
\end{equation}

Now for $\mathbf{\omega~||~y,}$ after averaging over $y,$ we get%
\begin{equation}
b_{x}=-k_{y}m_{z}\mu^{2}%
\end{equation}
and, if $\mathbf{\omega~||~x,}$ $b_{x}=0.$ We saw that there are some magnons
in the system that propagate along $y$ and can generate a topological magnetic
field $b_{x}$, which couples to the external field $H_{x}.$ By definition, the
energy cost to excite such a magnon is $Ak_{y}^{2}JM^{2},$ where $J$ is the
ferromagnetic exchange coupling in the plane. As we have seen, $b_{x}=Bk_{y}$
for magnons with some polarization planes and $-Bk_{y}m_{z}$ for others. The
coupling term must be $CH_{x}k_{y}.$ The constants $B$ and $C$ are
proportional to $m_{z},$ and are different for each plane. Let us calculate
the expectation value for the $\left\langle b_{x}\right\rangle $ for one plane
and one type of magnons:%
\begin{align}
\left\langle b_{x}\right\rangle  &  =\frac{\int Bk_{y}e^{-Ak_{y}^{2}%
J/T}(e^{CH_{x}k_{y}/T}-e^{-CH_{x}k_{y}/T})dk_{y}}{\int e^{-Ak_{y}^{2}%
J/T}(e^{CH_{x}k_{y}/T}+e^{-CH_{x}k_{y}/T})dk_{y}}\\
&  =\frac{\int Bk_{y}e^{-Ak_{y}^{2}J/T}\sinh(CH_{x}k_{y}/T)dk_{y}}{\int
e^{-Ak_{y}^{2}J/T}\cosh(CH_{x}k_{y}/T)dk_{y}}\\
&  \approx\frac{\int Bk_{y}e^{-Ak_{y}^{2}J/T}(CH_{x}k_{y}/T)dk_{y}}{\int
e^{-Ak_{y}^{2}J/T}dk_{y}}.
\end{align}
In the last line we made use of the fact that $J\gg T.$ From that,%
\begin{equation}
\left\langle b_{x}\right\rangle =BCH_{x}T/AJ=const\cdot TM_{z}^{2}%
H_{x}=const\cdot(1-M^{2}/M_{s}^{2})TH_{x},\label{THETH}%
\end{equation}
where $M_{s}$ is the saturated magnetization.

A requirement for this scenario is a conical spiral rotating in a plane
perpendicular to the external field. It is also essential that the coupling
between the planes is weak, allowing magnons to be excited independently in
each plane. It is also obvious from the general theory of the THE that
conduction electrons should be strongly coupled to the magnetic moments. This
implies that, as in MnSi \cite{Nagaosa2013s}, they belong to the same system
and have strong Hund's rule coupling. This is fulfilled here because both the
moments and the conductivity are due to Mn $d$-electrons, but may not work well
for, say, rare earth based spiral magnets. In any event, the scale of the
effect must be very material-dependent (coefficients $B$ and $C$ above), and
its microscopic evaluation may be challenging.\newline
\section{Topological Hall effect}\label{SI5}
The Hall effect, in general, is an intrinsic property of a conductor due to
the Lorentz force experienced by the charge carriers. In systems with
spontaneously broken time-reversal symmetry, an additional contribution,
independent of the Lorentz force, is observed which is proportional to the
magnetization $M$ and is called the anomalous Hall effect
(AHE)\cite{Nagaosa2010RMPs}. In materials with spin textures allowing a
non-zero scalar spin chirality defined by \textbf{\textit{S$_{i}$}} $\cdot$
(\textbf{\textit{S$_{j}$}} $\times$ \textbf{\textit{S$_{k}$}}), where $i,j,k$
are neighboring spins (equivalent, in the continuous approximation, to
Eq. \ref{N}), an additional component of the Hall effect is permitted due to
the real-space Berry phase called the topological Hall effect
(THE)\cite{Nagaosa2012s,Wang2019s}. Thus, a Hall resistivity can be expressed
as:
\begin{equation}
\rho_{H}=\rho^{O}+\rho^{A}+\rho^{T}.\label{E1}%
\end{equation}
Here $\rho^{O}=R_{0}B$ is the ordinary Hall resistivity, where $R_{0}$ is the
coefficient defined by the number of carriers (weighted with their mobility,
for a multiband metal), $B=\mu_{0}H$, and $H$ is the external magnetic field.
$\rho^{A}=R_{s}\mu_{0}M$ is the conventional anomalous Hall resistivity where
$R_{s}$ is the coefficient of the conventional AHE. $\rho^{T}$ is the Hall
resistivity contribution from the THE. $R_{S}$ and $R_{0}$ can be estimated
from the high magnetic field component of the magnetization and the Hall
resistivity (in the forced ferromagnetic state where the magnetization
saturates), where $\rho^{T}=0.$ Thus, Eqn. \ref{E1} takes the form:
\begin{equation}
\rho_{H}=R_{0}B+\mu_{0}R_{s}M.\label{E2}%
\end{equation}
The intercept of $\rho_{H}/M$ vs $B/M$ gives $\mu_{0}R_{s}$ while the slope
gives $R_{0}$. The $R_{0}$ estimated this way is correct in the forced
ferromagnetic (FF) state. However, we cannot assume the same in the low field
region where the Fermi surface (and hence the carrier concentration) is
different from that in the FF state. To address this discrepancy, we use the
following principle to estimate the normal component of the Hall resistivity
that needs to be subtracted (together with the anomalous Hall resistivity)
from the measured Hall resistivity to get the topological contribution.
Despite $M$ (and thus $\rho^{A})$ changing at $H_{1}$ nearly discontinuously,
up to a small spin-orbit coupling there is no discontinuity in the number of
carriers. At $H_{2},$ in principle, there might be a discontinuous change in
$R_{0},$ but since both TCS and FL phases at this point only slightly deviate
from the FF state, this change must be small. Thus, to a good accuracy, we can
assume that $R_{0}$ changes smoothly between the low-field regime $(H<H_{1}),$
where we can estimate it from the difference between $\rho_{H}$ and $\rho
^{A},$ and the high-field regime, where it is the only component changing. We
estimated the smooth change of the normal component of the Hall resistivity by
interpolating a cubic spline between $H_{1}$ and $H_{2}$ in the $\rho_{H}%
-\rho^{A}$ data. The measured $\rho_{H}$ and its different components are
shown in Fig. \ref{SFH1}(a). Together with $\rho^{O}$ estimated as explained
above (green solid line labelled 1), we also show, for comparison, $\rho^{O}$
obtained by using a simple linear interpolation between $H_{1}$ and $H_{2}$ in
the $\rho_{H}-\rho^{A}$ data (orange dashed line labelled 2), and $\rho^{O}$
calculated using $\rho^{O}=R_{0}B$, where $R_{0}$ is obtained in the FF state
(brown dashed line labelled 3). The THE obtained using \#1 and \#3 $\rho^{O}$
are depicted in Figs. \ref{SFH1}(c) and (d), respectively. The amplitude of
the THE obtained by using \#3 $\rho^{O}$ is slightly larger than that using
\#1 $\rho^{O}$, which is essentially due to an improper subtraction of the
normal Hall component in the former case as can be seen in Fig. \ref{SFH1}%
(d), where using \#3 $\rho^{O}$ still gives some THE contribution below
$H_{1}$, where it is not expected. The amplitude of the THE obtained using \#2
$\rho^{O}$ lies between that obtained with the other two $\rho^{O}$ (not shown).

In Fig. \ref{SFH1}(b), we show the calculated THE using Eqn. \ref{THETH} from
the theoretical model. As the THE is proportional to $\left\langle
b_{x}\right\rangle $, in the calculation we used:
\begin{equation}
\rho^{T}=\kappa(1-M^{2}/M_{s}^{2})TH\label{E4}%
\end{equation}
where $M$ is the magnetization measured in the magnetic field $H$, $M_{s}$ is
the saturated magnetization, $T$ is temperature, and $k$ is the
proportionality constant. In the calculation, the experimental magnetization
data measured at $T$ = 245 K are used (black solid line). It is to be noted
that Eqn. \ref{E4} is valid only in the TCS phase i.e. between $H_{1}$ and
$H_{2}$. Therefore, after calculating $\rho^{T}$ in the entire field range, we
determined the $\rho^{T}$ obtained outside the TCS phase as a background by
interpolating a straight line between $H_{1}$ and $H_{2}$ (dashed pink line)
and subtracted the background to obtain the THE in the TCS phase (solid blue
line). This theoretical $\rho^{T}$ is compared to the experimental data in
Fig. \ref{SFH1}(c), which is also shown in Fig. 4(d). In Fig. \ref{SFH1}(d),
we compare the theoretical THE with the experimental THE obtained by using \#3
$\rho^{O}$ discussed above. From Figs. \ref{SFH1}(c) and (d), we see that the
theoretical model describes the experimental data fairly well irrespective of the method used to estimate 
the normal Hall component (which is much smaller than the THE). The only
difference in the calculated THE in these two cases is the proportionality
constant $\kappa$.

To provide further evidence of the THE in YMn$_{6}$Sn$_{6}$ for the in-plane
magnetic field, we show the Hall resistivity measured with the magnetic field
in the $ab$-plane and along the $c$-axis in Fig. \ref{SFH2}(a). This shows
that a topological Hall contribution appears between around 2 T and 5 T only
in the case when the magnetic field is applied in the $ab$-plane. The
corresponding magnetization data for comparison are presented in Fig. \ref{Fig2}(a).
The Hall resistivity measured with magnetic field applied in the $ab$-plane at
5 K and 245 K is presented in Fig. \ref{SFH2}(b) and the corresponding
magnetization data are presented in Fig. \ref{SFH2}(c). The difference in
magnetization between 5 K and 245 K below 6 T is small, but the Hall
resistivity at 245 K is highly enhanced as compared to that at 5 K, which also
supports the presence of a topological contribution to the Hall resistivity at
245 K.

\section{Supplementary References}
\bibliographystyle{naturemag}

\newpage
\begin{table}[th]
\caption{Calculated exchange and single-ion energies for the ``Full", ``reduced", and ``minimal" models.}%
\label{T1}
\begin{center}
\par%
\begin{tabular}
[c]{lccc}\hline
& Full model & Reduced Model & Minimal model\\\hline
$J_{1}$ & $-12.86$ & $-12.86$ & $-12.86$\\
$J_{2}$ & $7.26$ & $4.66$ & $4.66$\\
$J_{3}$ & $-0.06$ & $-2.20$ & $-2.20$\\
$J_{4}$ & $0.06$ & - & -\\
$J_{5}$ & $-0.16$ & - & -\\
$J_{6}$ & $0.54$ & - & -\\
&  &  & \\
$J_{p}$ & $-53$ & $-53$ & $-53$\\
$K$ & $-0.31$ & $-0.31$ & $0.19$\\
$J^{z}$ & $0.50$ & $0.50$ & -\\\hline
\end{tabular}
\end{center}
\end{table}

\begin{figure}[th]
\begin{center}
\includegraphics[scale=.84]{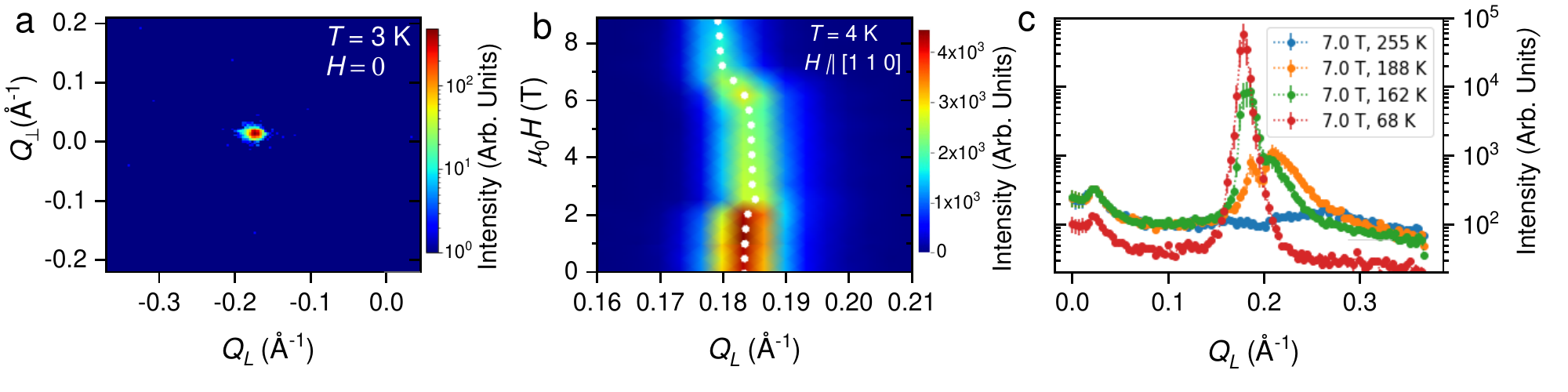}
\end{center}
\caption{ {\textbf{Small angle neutron scattering (SANS)) of YMn$_6$Sn$_6$ .} }(a) Background-subtracted SANS image on the two-dimensional position sensitive detector at 4 K and without a magnetic field applied.  The data show a single incommensurate magnetic peak at a wave vector of $Q$ 
= 0.184 Å$^{-1}$, corresponding to a real-space modulation of 34 Å.  No evidence for a skyrmion lattice was found at any temperature or field. (b) Intensity-$Q$ map of the field dependence of the Bragg peak at 4 K, revealing a series of phase transitions. The white dots are a guide to the peak center. (c) ­­­Cuts of the observed intensity through the magnetic peak at 7 T and for a series of temperatures, where it was discovered that at higher temperatures there are two closely-spaced incommensurate peaks.  Subsequent wide-angle high resolution diffraction data revealed that there are two wave vectors at all temperatures. Note that the apparent peak at very small $Q$ is simply due to an incomplete subtraction around the beam stop.} \label{SFN6}%
\end{figure}

\begin{figure}[th]
\begin{center}
\includegraphics[scale=.75]{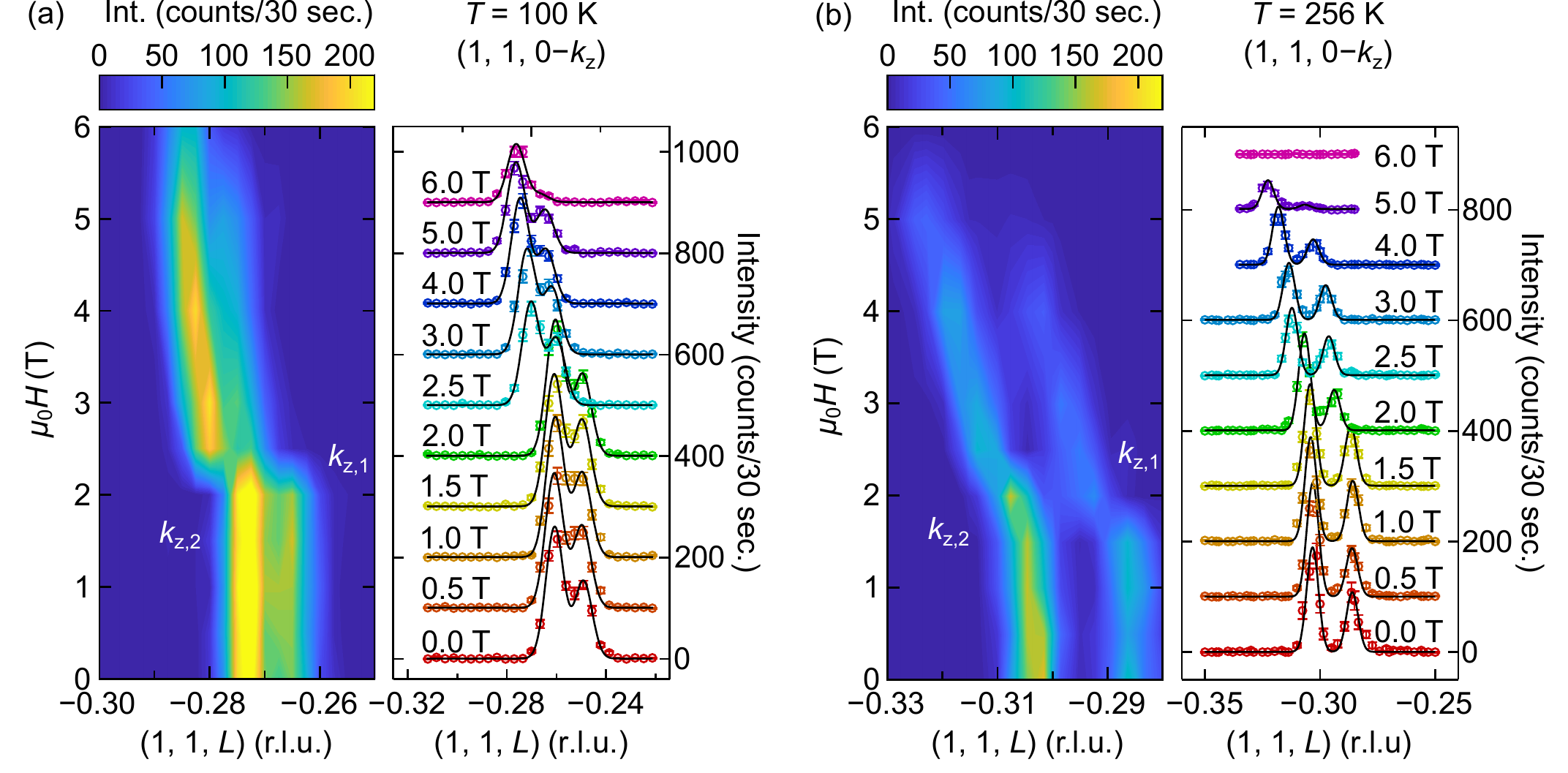}
\end{center}
\caption{ {\textbf{Temperature and magnetic field dependence of incommensurate
magnetic Bragg peaks (1, 1, $0-\bm{k_z}$).} }a) The evolution of Bragg peaks
$k_{z,1}$ and $k_{z,2}$ with an applied magnetic field at 100K and b) 256 K. For both (a) and (b), the solid black lines in the right-hand panels are Gaussian fits to the data as described in the text. An offset of 100 counts/30 sec. was added between individual $L$ scans for clarity. } \label{SFN1}%
\end{figure}

\begin{figure}[th]
\begin{center}
\includegraphics[scale=.5]{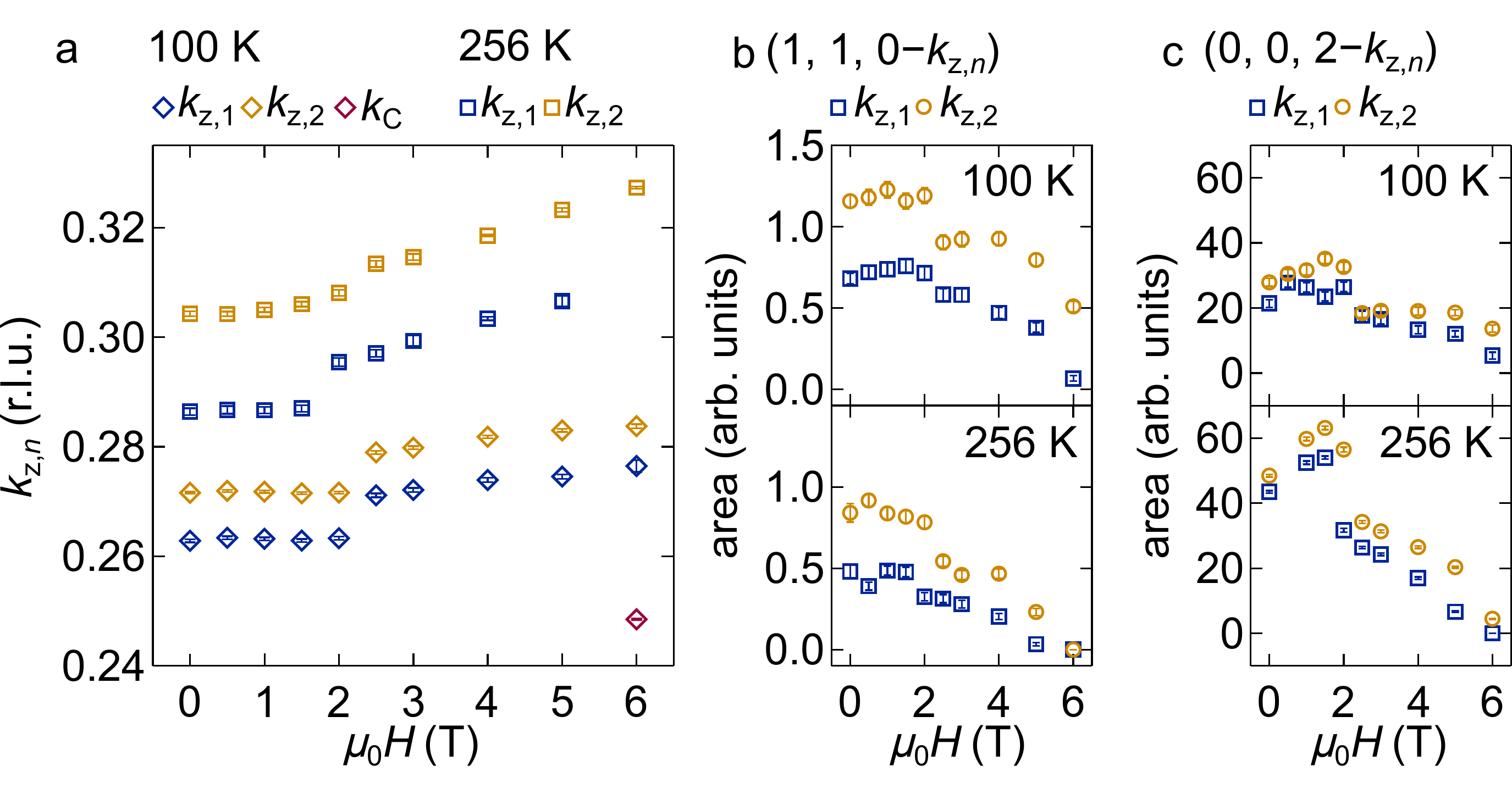}
\end{center}
\caption{\textbf{Temperature and magnetic field dependence of wave vector and
intensity of incommensurate magnetic Bragg peaks.} a) The evolution of wave vectors (0, 0, $k_{z,1}$) and (0, 0, $k_{z,2}$) depicted in Fig. 3(a) of the
main text with applied magnetic field at 100 K and 256 K. An additional
commensurate peak, $k_{c}$ = 0.25, appears at 6 T in the 100 K data. b)
Integrated intensity of the Bragg peaks (1, 1, $0-k_{z,n})$ depicted in
Fig. \ref{SFN1} at 100 K and 256 K. c) Integrated intensity of
Bragg peaks (0, 0, $2-k_{z,n})$ depicted in Figs. 3(b) and (c) of the main
text as a function of magnetic field applied along [1,
$\bar{1}$, 0].}%
\label{SFN2}%
\end{figure}

\begin{figure}[th]
\begin{center}
\includegraphics[scale=.5]{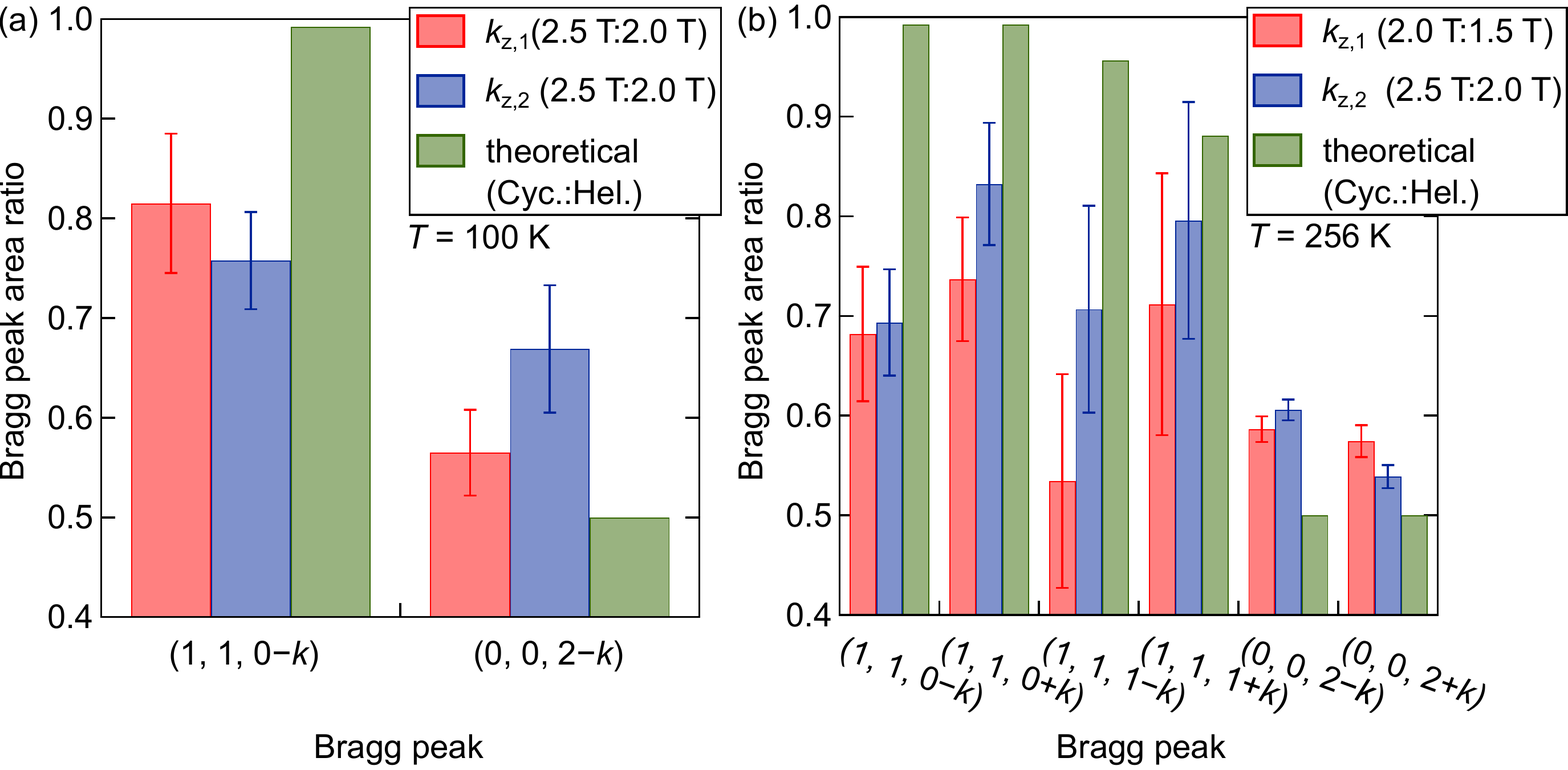}
\end{center}
\caption{{\textbf{Comparison of cycloidal vs helical spiral intensity.} Ratio
of intensities just above $H_{1}$ to just below $H_{1}$ for the incommensurate
Bragg peaks measured at (a) 100 K and (b) 256 K. The calculated values of the
cycloidal:helical magnetic structure factor ratios are shown in green for
comparison.}}%
\label{SFN3}%
\end{figure}

\begin{figure}[th]
\begin{center}
\includegraphics[scale=.2]{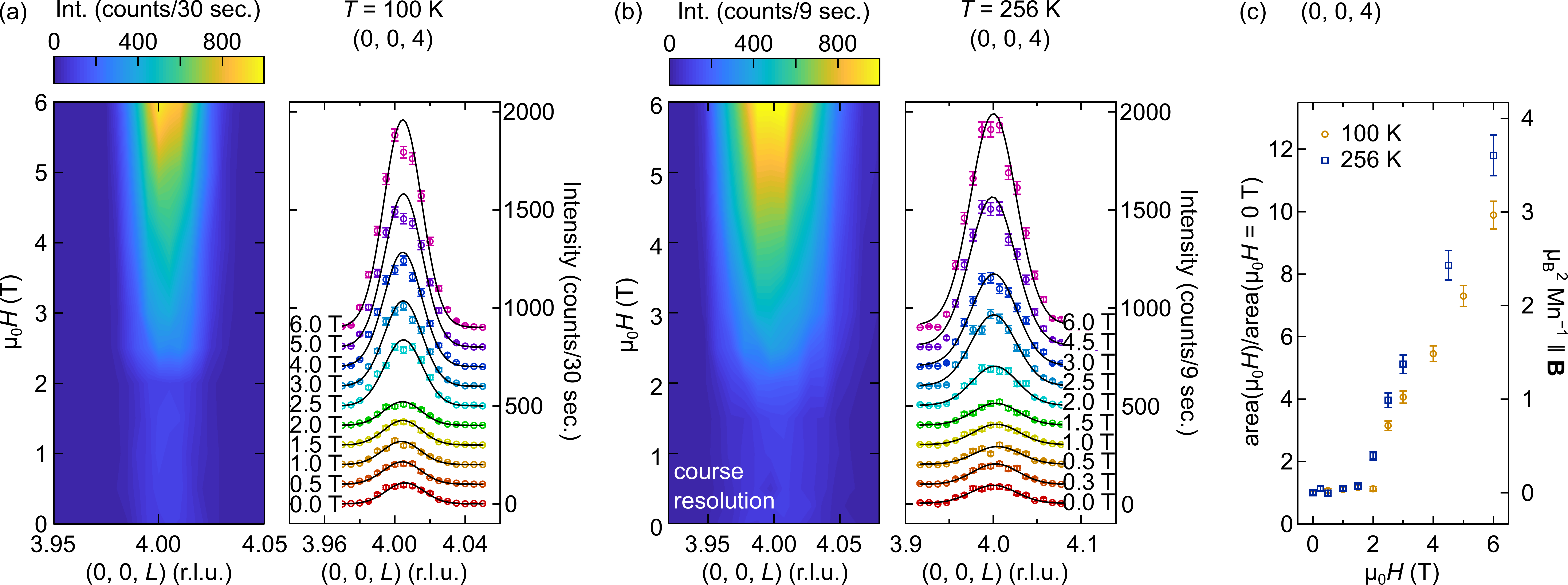}
\end{center}
\caption{{\textbf{Nuclear Bragg peak intensity as a function of magnetic
field.} (0, 0, 4) Bragg peak intensity tracked at (a) 100 K and (b) 256 K as a
function of applied magnetic field applied along [1, $\bar{1}, 0]$
crystallographic axis. The increase in intensity as the field increases is
indicative of a net component of magnetization emerging due to the moments
canting towards the applied field direction. The solid lines in the
righthand panels of (a) and (b) are the Gaussian fits to the data described in
the text. An offset was added between individual $L$ scans for clarity
(offsets are 100 counts/30 sec. for (a) and 100 counts/9 sec. for (b)). Panel
(c) shows the intensity versus $H$, divided by the intensity at $H$ = 0 where
only the nuclear structure contributes intensity. The right hand axis displays
the projection of the moment, in $\mu_{B}^{2}$Mn$^{-1}$, along the applied
field direction.}}%
\label{SFN4}%
\end{figure}

{\begin{figure}[th]
\begin{center}
\includegraphics[scale=1]{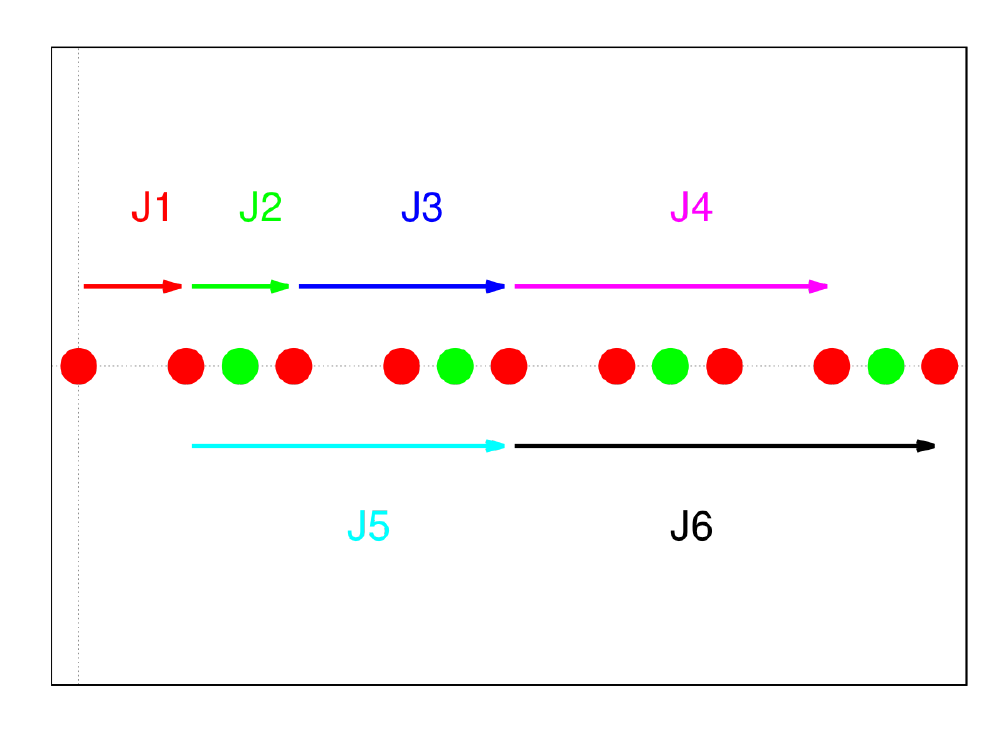}
\end{center}
\caption{\textbf{Schematic of exchange interactions along $c$-axis.} First 6
exchange interactions between Mn layers. Red: Mn layer; green: spacer layer
including Y [see Fig. \ref{Fig1}(a) of the main text]}%
\label{FigJ}%
\end{figure}}

{\begin{figure}[th]
\begin{center}
\includegraphics[scale=1.5]{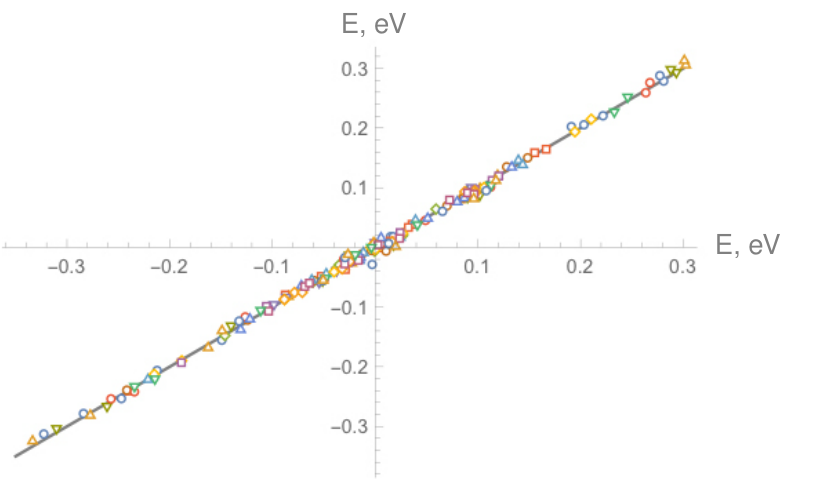}
\end{center}
\caption{Fitted vs. calculated values of total energies (deviations from
the straight line indicate fitting errors). Blue circles are the calculated
values for $U-J=0$. Other symbols are for other values of $U$.}%
\label{fitquality}%
\end{figure}}

{\begin{figure}[th]
\begin{center}
\includegraphics[scale=1.5]{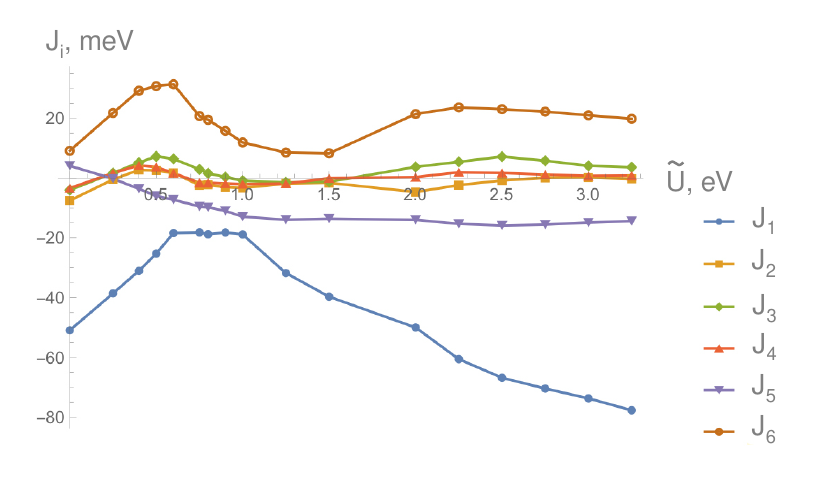}
\end{center}
\caption{ Fitted values of the exchange constants $J_{1}$--$J_{6}$ as a
function of $\tilde{U}=U-J$.}%
\label{UJ}%
\end{figure}}

{\begin{figure}[th]
\begin{center}
\includegraphics[scale=1]{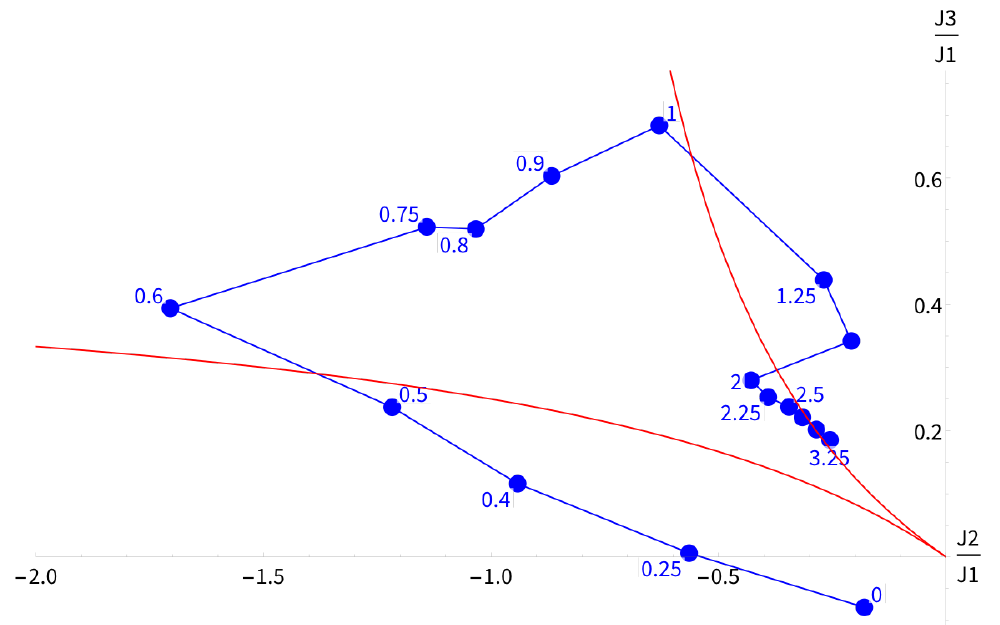}
\end{center}
\caption{Phase diagram of the $J_{1}-J_{2}-J_{3}$ model in the $J_{2}%
/J_{1}-J_{3}/J_{1}$ coordinates. The region between the two red lines is a
spiral state, above them is the FM, and below the AF $uudd$ state. The points
reflect the calculated values of $J_{1},J_{2},J_{3}$ for different values of
$U-J$.}%
\label{GS}%
\end{figure}

{\begin{figure}[th]
\begin{center}
\includegraphics[scale=1]{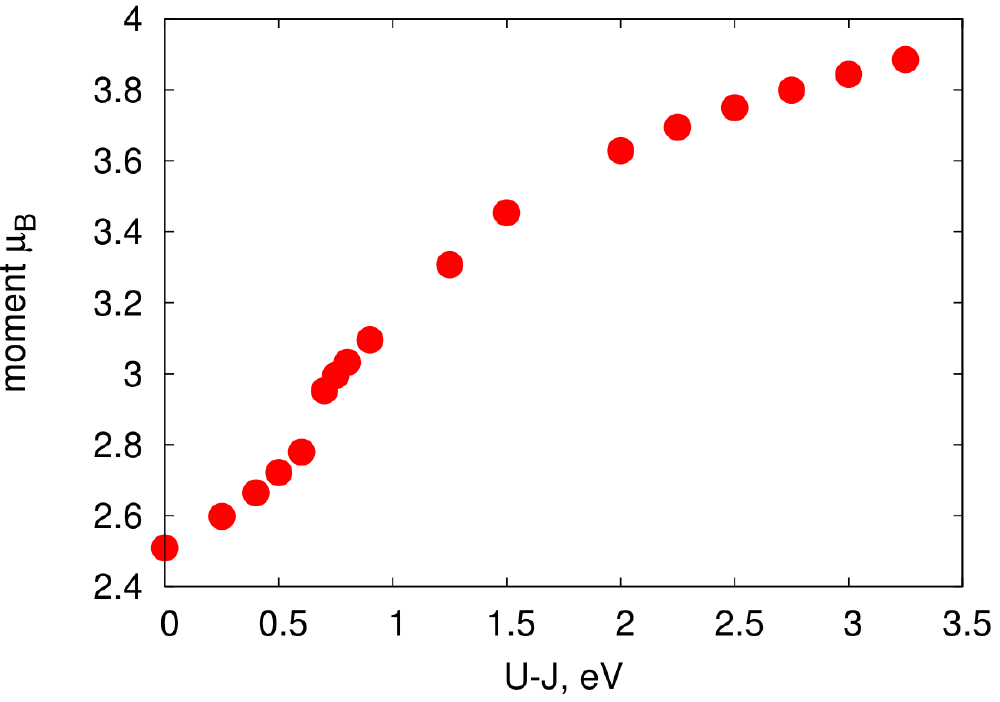}
\end{center}
\caption{Calculated magnetic moments on Mn (square-averaged over all 8 sites
and all 10 configurations) as a function of $U-J$.}%
\label{M}%
\end{figure}}

{\begin{figure}[th]
\begin{center}
\includegraphics[scale=1.2]{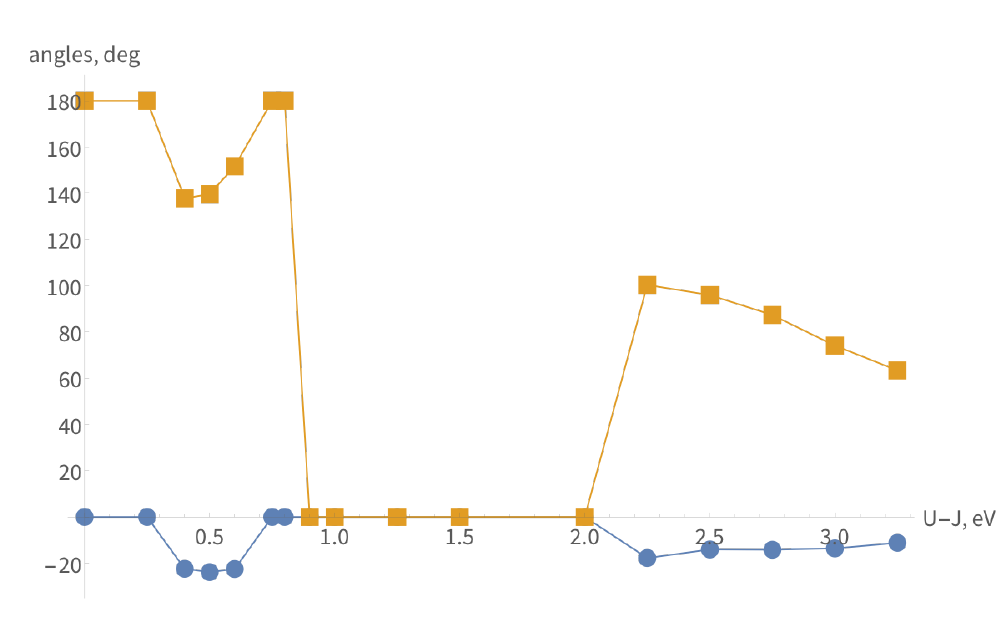}
\end{center}
\caption{Calculated spiral angles $\alpha$ and $\beta$ (see the text for the
definitions).}%
\label{angles}%
\end{figure}}

{\begin{figure}[th]
\begin{center}
\includegraphics[scale=1]{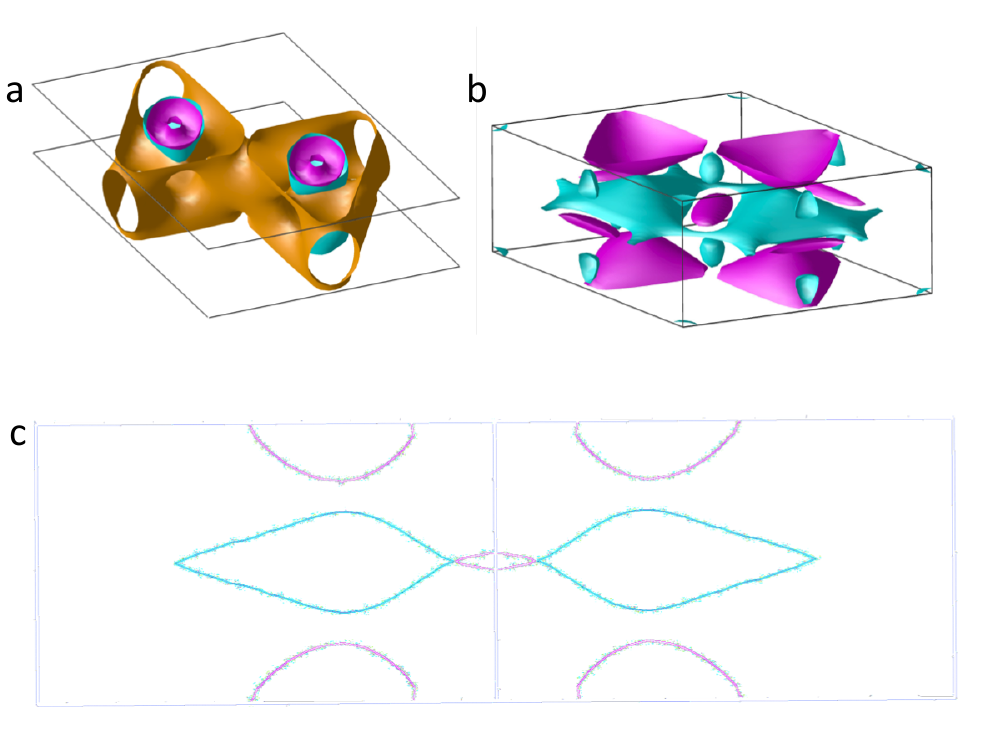}
\end{center}
\caption{Calculate Fermi surface (no Hubbard $U$) for the ferromagnetic
ordering, for a) the spin-minority and b) spin-majority electrons. c) A
vertical cut of the spin-majority Fermi surface in the $\Gamma$-K-K-A plane.}%
\label{FS}%
\end{figure}}

\begin{figure}[th]
\begin{center}
\includegraphics[scale=.7]{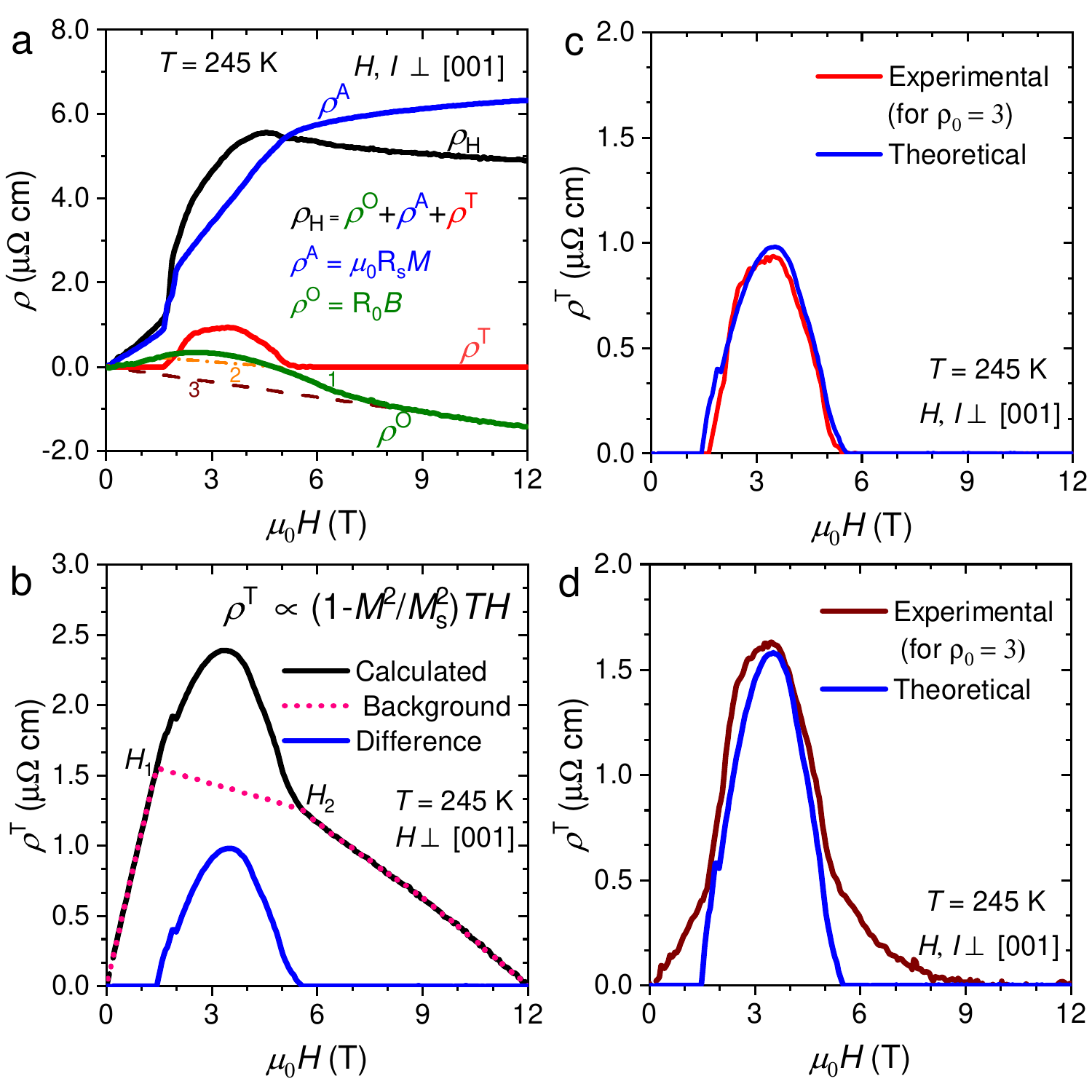}
\end{center}
\caption{ \textbf{Estimation of topological Hall resistivity at 245 K from measured data and theoretical model.} a) Hall
resistivity and its various components. Three different $\rho^{0}$  are labelled as 1, 2 and 3 and are discussed in the text in \ref{SI5}. b) Topological Hall effect calculated using the chiral spin texture model discussed in \ref{SI4}. Since there is no THE outside of TCS phase, the component obtained outside this phase is subtracted as a background  (pink dashed line) from the calculation carried out in the entire field range (black solid line) from 0 to 12 T to obtain the THE (blue solid line)  in the TCS phase. c) Topological Hall resistivity estimated using $\rho^{0}$ labelled as 1 in panel (a) compared to that obtained form the theoretical model. d) Topological Hall resistivity estimated using $\rho^{0}$ labelled as 3 in panel (a) compared to that obtained form the theoretical model. Results presented in panels (c) and (d) show that the theoretical model describes the experimental data well irrespective of the method used to estimate the ordinary component of the Hall resistivity ($\rho^{0}$).}%
\label{SFH1}%
\end{figure}

\begin{figure}[th]
\begin{center}
\includegraphics[scale=1]{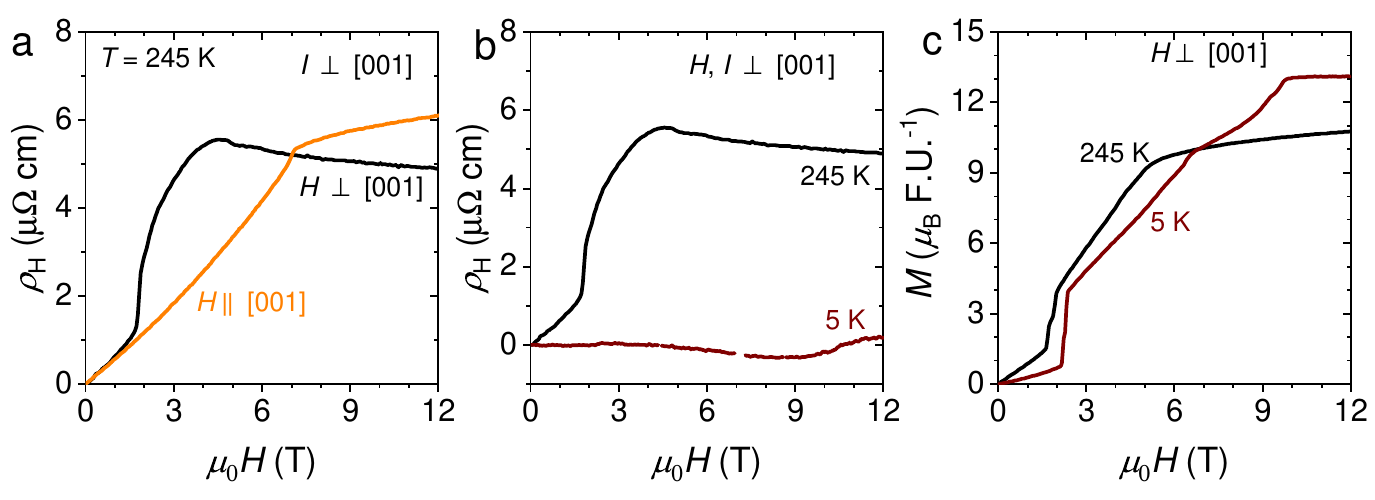}
\end{center}
\caption{ \textbf{Hall resistivity and magnetization of YMn$_6$Sn$_6$.}  a) Hall resistivity
at 245 K measured with magnetic field in the $ab$-plane and along the
$c$-axis. The current was applied along the same direction in these two measurements
carried out on two different samples. b) Hall resistivity at 5 K and 245 K
measured with magnetic field applied in the $ab$-plane. c) Magnetization at 5 K
and 245 K measured with the magnetic field in the $ab$-plane.}%
\label{SFH2}%
\end{figure}
\begin{figure}[th]
\begin{center}
\includegraphics[scale=.8]{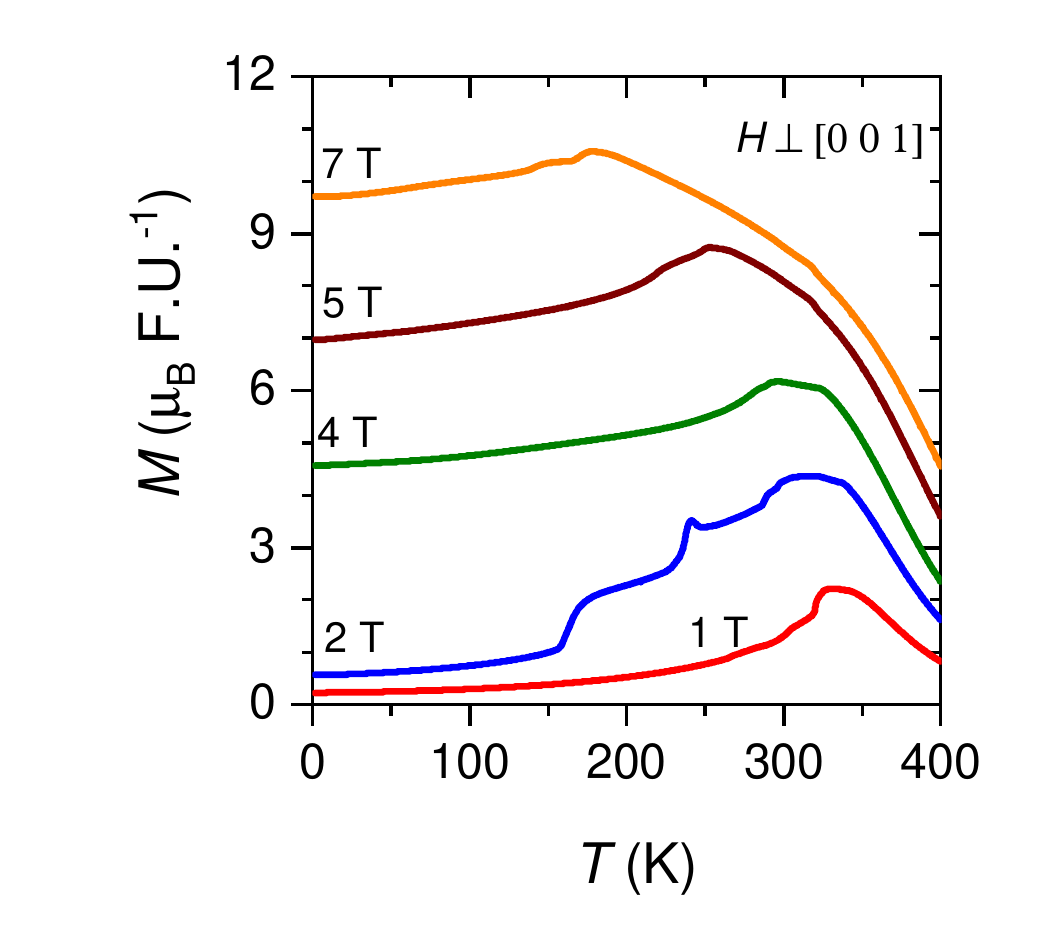}
\end{center}
\caption{ \textbf{Magnetic moment of YMn$_6$Sn$_6$ as a function of temperature.} Magnetic moment as a function of temperature at indicated magnetic fields applied in the $ab$-plane. Above 3 T and below 250 K, magnetic moment changes very little with the temperature.}%
\label{SFH3}%
\end{figure}

\end{document}